\documentclass[fleqn,usenatbib]{mnras}

\usepackage{newtxtext,newtxmath}
\usepackage[T1]{fontenc}
\usepackage{ae,aecompl}

\usepackage{graphicx}	% Including figure files
\usepackage{amsmath}	% Advanced maths commands
\usepackage{amssymb}	% Extra maths symbols
\usepackage{verbatim}
\usepackage{xcolor}

\title{Support Vector Machine classification of strong gravitational lenses}

\author[Hartley et al.]{
P. Hartley,$^{1}$\thanks{E-mail: philippa.hartley@postgrad.man.ac.uk}
R. Flamary,$^{2}$
N. Jackson$^{1}$
A. S. Tagore$^{1}$
and R. B. Metcalf$^{3}$
\\
$^{1}$Jodrell Bank Centre for Astrophysics, School of Physics \& Astronomy, University of Manchester, Oxford Rd., Manchester M13 9PL, UK\\
$^{2}$Laboratoire Lagrange, Universit\'e C\^ote d'Azur, CNRS, Observatoire de la C\^ote
d'Azur, Parc Valrose, 06108 Nice, France\\
$^{3}$Department of Physics and Astronomy, Universit\`a di Bologna, viale Berti Pichat 6/2, 40127 Bologna, Italy
}

\date{Accepted XXX. Received YYY; in original form ZZZ}

\pubyear{2017}

\begin{document}
\label{firstpage}
\pagerange{\pageref{firstpage}--\pageref{lastpage}}
\maketitle

\begin{abstract}
The imminent advent of very large-scale optical sky surveys, such as Euclid and LSST, makes it important to find efficient ways of discovering rare objects such as strong gravitational lens systems, where a background object is multiply gravitationally imaged by a foreground mass. As well as finding the lens systems, it is important to reject false positives due to intrinsic structure in galaxies, and much work is in progress with machine learning algorithms such as neural networks in order to achieve both these aims. We present and discuss a Support Vector Machine (SVM) algorithm which makes use of a Gabor filterbank in order to provide learning criteria for separation of lenses and non-lenses, and demonstrate using blind challenges that under certain circumstances it is a particularly efficient algorithm for rejecting false positives. We compare the SVM engine with a large-scale human examination of 100000 simulated lenses in a challenge dataset, and also apply the SVM method to survey images from the Kilo-Degree Survey.
\end{abstract}

\begin{keywords}
gravitational lensing: strong -- galaxies: general -- methods: statistical -- methods: data analysis -- surveys
\end{keywords}

%%%%%%%%%%%%%%%%%%%%%%%%%%%%%%%%%%%%%%%%%%%%%%%%%%

%%%%%%%%%%%%%%%%% BODY OF PAPER %%%%%%%%%%%%%%%%%%

%\large   % Or I have to use my glasses to read it!

\section{Introduction}

The discovery of the first strong lens system, the twin object QSO 0957+561 \citep{1979Natur.279..381W}, heralded a new line of scientific inquiry. Strong lenses are systems in which the light from a background object is deflected by a foreground galaxy or group of galaxies, resulting in multiple images of the background object. Study of the images gives unique information about the distribution of matter in the deflector, independent of the light it emits. This allows us to constrain galaxy mass models (e.g. \citealt{2002ApJ...575...87T,2003MNRAS.343L..29T,2004ApJ...611..739T,0004-637X-752-2-163,0004-637X-777-2-98,0004-637X-800-2-94}), dark matter substructure (e.g. \citealt{maosch,2004ApJ...610...69K,2012Natur.481..341V,2001ApJ...563....9M}) and, via the lensing of time-varying objects, the Hubble constant (e.g. \citealt{1964MNRAS.128..307R,2004mmu..symp..117K,2014ApJ...788L..35S,2016arXiv161100014G}). The magnification effect of lensing also allows us to use strong lenses as cosmic telescopes, enabling us to observe otherwise unseen objects such as QSO host galaxies and radio quiet quasars (e.g. \citealt{2006A&A...451..865C,2011ApJ...739L..28J,2015MNRAS.454..287J}).

Lens systems are rare objects; typically, for every 1000--2000 objects at high redshift, one will be multiply imaged by a foreground deflector\footnote{The exact number depends on details of the source population; populations with steep number counts, noticeably at submillimetre wavelengths, have much higher lensing rates (e.g. \citealt{1996MNRAS.283.1340B}).}. Nevertheless, since the first lens system was found, several hundred more have been discovered, and much work has been done to use this population to convert the theoretical scientific potential into results (see \citealt{2010ARA&A..48...87T} for the most recent comprehensive review). However, while the number of known lens systems is increasing, many scientific applications require special properties seen only in subsets of the sample. For example, the use of lenses for cosmography, and in particular the determination of the Hubble constant $H_0$, requires a lens system with a variable source { \citep{1964MNRAS.128..295R}, together with an extended distribution of lensed flux in the image plane and, ideally, high-quality observations of stellar kinematics
%several important subsets within the sample remain sparse. 
%Time-varying lensed sources .. more detail about $H_0$ %requirements, velocity dispersions...
\citep{2016arXiv160700017S,2017MNRAS.465.4895W,2017MNRAS.465.4914B}. Constraints on the dark-energy equation of state parameter, $w$, require double-source plane lens systems  \citep{2012MNRAS.424.2864C,2014MNRAS.443..969C}, which occur still more rarely (once for every few hundred ``normal'' lens systems). In order to make progress, therefore, increases of orders of magnitude in the sizes of lens samples are required.
%Suyu H0liCOW including systems with time-varying lensed sources and those with well-modelled lensing components, remain sparse.} Furthermore, cosmological measurements that require large samples of systems are not well-constrained using the current sample. \rf{RF: paragraph unclear. Last point evry interesting but reference needed.}

The advent of a new era of astronomical observation will provide samples of lens systems orders of magnitudes larger than those which already exist. The Square Kilometre Array (SKA) \citep{2011arXiv1105.5953R}, the Large Synoptic Survey Telescope (LSST) \citep{2009arXiv0912.0201L} and the Euclid telescope \citep{2011arXiv1110.3193L} will all see first light within the next decade. These new surveys, designed to image large areas of the sky with high sensitivity, will each observe billions of sources, yielding a statistically robust sample of strong lenses for the first time \citep{2015aska.confE..84M,2010MNRAS.405.2579O}. 

Traditionally, detection of these rare objects has relied upon visual inspection of survey data.  The Cosmic Lens All Sky Survey (CLASS) \citep{2003MNRAS.341...13B} found 22 confirmed lensed quasar systems from an initial source sample of 16503. The sample was reduced using a combination of repeated visual inspection by several team members along with a figure of merit obtained by modelling each source. More recently, visual inspections of the HST legacy programme investigations of the COSMOS field have been used to find lens systems, first by \cite{2008ApJS..176...19F} in a sample of $\sim$7000 objects. Subsequently, the whole COSMOS object sample of about 280000 sources in the COSMOS field was reduced to 112 lens candidates and two new confirmed lenses by eye alone \citep{2008MNRAS.389.1311J}. Although successful, visual inspection by a small team is time-consuming, and becomes unfeasible in the case of future, much larger, surveys, where billions of sources will be imaged. 

One approach to address the time consuming problem of the lens-finding process has led to the SPACE WARPS project \citep{2015arXiv150406148M,2015arXiv150405587M}. Developed by the creators of the source-classification project Galaxy Zoo \citep{2013MNRAS.435.2835W}, SPACE WARPS uses crowdsourcing via an online platform in order to identify strong lens systems from legacy survey maps. To date, a few tens of new candidates have been identified \citep{2016MNRAS.455.1191M} from large ground-based surveys such as the Canada-France-Hawaii Telescope Legacy Survey\footnote{\tt {http://www.cfht.hawaii.edi/science/cfhtls}}. However, the number of classifications made is still orders of magnitude below that required for lens detection in the large surveys of the future.  At least some degree of automation is necessary if this goal is to be met. It is also unknown to what degree classification by the human eye is subject to bias against, for example, small-separation lens systems. It would be useful to determine this bias and discover whether sub-populations of lens systems are being missed.

Static computer algorithms have so far had some success in identifying lens candidates, both in simulations and from real data. \cite{2009ApJ...694..924M} devised a lens-modelling  `robot' that can be applied to Bright Red Galaxies (BRGs) to determine whether a lensed source is present. The PCA lens-finder \citep{2014A&A...566A..63J} first uses principal component analysis to remove the lensing galaxy before searching for lensed features using an island detection algorithm. Application of the PCA finder to the Canada France Hawaii Telescope Legacy Survey (CFHTLS) found 109 new lens candidates. Ringfinder  uses difference imaging to detect blue residuals embedded in otherwise smooth red light distributions surrounding early type galaxies  \citep{0004-637X-785-2-144}. Arcfinder \citep{2007A&A...472..341S} uses a pixel-grouping technique to find cluster-scale lensing events. The methods are informed by scientific expertise - in this case that of the lensing phenomenon - and judicious incorporation of this knowledge has so far shown promise with application to real datasets \citep{2012ApJ...749...38M,2016A&A...592A..75P}. However, prior knowledge could also lead to bias in techniques; if the lensing populations are not fully understood, then such approaches may miss potentially interesting lens system objects. 

Recent advances in artificial intelligence (AI) have permitted a new generation of approaches to image recognition. At the time of writing, a considerable number of applications of various types of AI to astronomical images have been published, and more are currently being developed. The application of convolutional neural networks (CNN) to galaxy classification has shown good results \citep{2015MNRAS.450.1441D}. CNN have since been used  \citep{2017arXiv170207675P} to find 56 lens candidates in the Kilo Degree Survey \citep{2013ExA....35...25D} and the use of Gaussian mixture modelling \citep{2016arXiv160701391O} has been used to discover a new lens system from multi-survey observations. Support vector machines (SVM) have been applied to several problems within astronomy. \cite{6643007} and \cite{2008A&A...478..971H,2009A&A...497..743H} have used SVM algorithms to classify galaxy morphologies, \cite{e484a2b0b8d84de4b3e1f2415f602fd1} applied the technique to supernova recognition and \cite{MNR:MNR21191} have used SVM to develop a quasar candidate classification system. Previous successful applications of SVM to the lens finding problem are lacking in the literature. 

The AI approach is often a general one, particularly in the case of CNN, where the algorithm can itself learn which features within input data are important, with no need for prior scientific information. By comparing results from human lens detection with those from AI from the same, large, dataset, this generality can help to inform the human of lens system characteristics which have previously not been considered. The SVM method we have developed is similarly general in its approach, with a feature set obtained by the use of a Gabor filtering kernel, a popular choice within machine learning image recognition methods and thought to mimic part of the image processing functions of the mammalian brain \citep{Jones1233}. While our method is more general than the static algorithm approaches, it may still suffer from problems of bias; with only a relatively small sample of observed lens systems available, training of the algorithm relies on simulation images, and will therefore be dependent on the quality of the model used to derive the training dataset. In section 2 we present the SVM method, and describe its implementation and lens-finding performance on test data sets. In section 3 we describe its performance on a recent public data challenge, and compare this with manual inspection. In section 4, we describe its application to recent observational surveys, before presenting the conclusions and suggestions for further development in section 5.

\section{SVM  implementation}

A SVM is a supervised machine learning method, and as such requires a set of labelled training data - a training set - in order to learn a model. The model can then be used to make predictions for new data. Provided that the new data are drawn from the same distribution and that the classifier is properly regularised, an SVM can make accurate predictions on new data, \emph{i.e.} it will generalise. In the case of lens-finding, the problem is a binary one involving two classes: images containing a lensed source and images without a lensed source. In galaxy-galaxy lensing, extended lensed sources typically appear as arc-like objects and  sometimes complete rings, while unresolved sources such as quasar cores appear as multiple point sources. Our methods were developed using the python scikit-learn and scikit-image packages \citep{scikit-learn,scikit-image}, which contain machine learning and image processing libraries. The finder was trained and tested using 4 3.3GHz CPUs.

\subsection{SVM}

SVM classification is strongly rooted in statistical theory. Originally developed by \citeauthor{vapnik79estimation} in 1979, the idea is to represent each training image sample as a vector $\mathbf{x}_i$ in multi-dimensional feature space. Associated with each image is a label $y_i$ representing a class. In the case of lens-finding, the problem is a binary one and we define the labels $y = +1$ and $y = -1$ for images which do and do not contain lensed sources, respectively. We can then select a hyperplane with which to separate classes (Fig.~\ref{svm}). Defining a weight vector $\mathbf{w}$ normal to the hyperplane and a bias term $b$ so that  $-b/\lVert\mathbf{w}\rVert$ is the perpendicular distance from the origin, all points on the hyperplane must satisfy the condition  $\mathbf{w}\cdot\mathbf{x}+b = 0$. We could solve the problem using one of many possible separating hyperplanes, but we would like to find the hyperplane which provides the best distinction between classes in order to minimise generalisation error when applying the model to unseen data. To do this we maximise the margin between the hyperplane and each class. By ensuring that all training samples satisfy the following constraints:

\begin{equation}
\begin{split}
&\mathbf{w}\cdot\mathbf{x}_i  +b \geq +1 \:\:\:\mathrm{for}\: y_i = +1\:\:\: \mathrm{and} \\
&\mathbf{w}\cdot \mathbf{x}_i +b \leq -1 \:\:\:\mathrm{for}\: y_i = -1,
\end{split}
\end{equation}

we can identify so-called support vectors at the inner edge of each class which meet conditions  $\mathbf{w}\cdot\mathbf{x}+b = +1$ or $\mathbf{w}\cdot\mathbf{x}+b = -1$. Rearranging these conditions using the distance from the origin, the size of the margin is found to be $2/\lVert\mathbf{w}\rVert$ and can be maximised by selecting support vectors which minimise the term $\lVert\mathbf{w}\rVert^2$.

The optimisation is formulated in the Lagrangian

\begin{equation}
L_P \equiv \frac{1}{2}\lVert\mathbf{w}\rVert^2 -\sum_{i=1}^N \alpha_i y_i (\mathbf{x}_i \cdot \mathbf{w} + b) + \sum_{i=1}^N \alpha_i
\end{equation}

using Lagrange multipliers $\alpha_i$ for each of $N$ training samples. Setting the gradient $\nabla{L}_P$ to zero, we obtain solutions
\begin{equation}
\mathbf{w} = \sum_i \alpha_i y_i \mathbf{x}_i
\label{w}
\end{equation}

and

\begin{equation}
\sum_i \alpha_i y_i = 0,
\end{equation}

allowing us to reformulate the problem in the simpler form: 

\begin{equation}
L_D = \sum_i \alpha_i - \frac{1}{2} \sum_{i,j} \alpha_i \alpha_j y_i y_j \mathbf{x}_i\cdot \mathbf{x}_j.
\label{Lagr2}
\end{equation}

Lagrange multipliers $\alpha_i$ ultimately form weighting coefficients representing the contribution of each training sample $\mathbf{x}_i$ to the solution; the support vectors are those samples where $\alpha_i >0$. From equation~(\ref{Lagr2}) it can be seen that optimisation depends only on the dot-products, $\bf{x}_i\cdot\bf{x}_j$,  of support vectors, resulting in a computationally lightweight process even for very large data dimensionality. In the non-separable case soft margins can be used at the expense of some misclassifications \citep{Cortes1995}. The optimisation process is convex, producing a single, global solution and avoiding the problem of local minima suffered by neural networks \citep{Burges1998}. After training, a prediction $\hat{y}$ for an unseen data sample $\mathbf{x}'$ can be made by evaluating $\mathbf{w}\cdot\mathbf{x}'+b$. Using the solution from equation~(\ref{w}), our final classifier can be expressed as

\begin{equation}
\hat{y}=  \sum_i \alpha_i y_i (\mathbf{x}_i\cdot \mathbf{x}')+b.
\label{classifier}
\end{equation}

\begin{figure}
  \centering
  \includegraphics[width=1\linewidth]{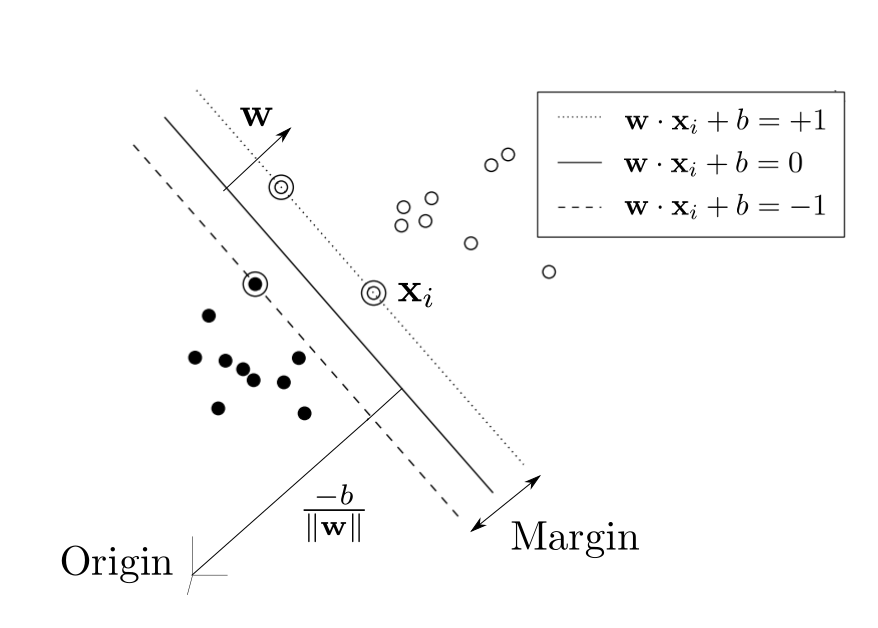}\\
   \caption{The SVM projects two classes (filled and unfilled circles) of training data samples as vectors $\mathbf{x}_{i,...,N}$ into multidimensional space, before defining a separating hyperplane. The solution is optimised by maximising the margin between the hyperplane and a set of support vectors (circled) on the inner edge of each class. The problem is formulated using weight vector $\mathbf{w}$, bias $b$  and class labels $y=+1$ and $y=-1$, such that a Lagrangian can be used to maximise the margin. Optimisation depends ultimately on only the dot products of the support vectors.}
  \label{svm}
  \end{figure}

Further development of the method \citep{Boser:1992:TAO:130385.130401} showed that if a coordinate transformation $\phi$ is applied to the input vectors, then SVM could be used to solve highly non-linear problems (Fig.~\ref{coord}). Better still, \cite{aizerman1964theoretical} showed that the specific coordinate transformation required for successful classification does not need to be known. Using a kernel function corresponding to the feature space  \citep{Mercer415}, and exploiting the method developed by \cite{aronszajn1950theory} to represent kernels in linear spaces, the dot-product dependence of the SVM permits the use of kernel function $K$, which depends only on the dot product of the transformed input vectors:
\begin{equation}
K(\mathbf{x}_i,\mathbf{x}_j) = \phi(\mathbf{x}_i)\cdot\phi(\mathbf{x}_j).
\end{equation}
This so-called `kernel trick' can be used to train and test data in higher dimensional space, in a similar amount of time as would be taken in linear space, without the need to explicitly know the coordinate transformation involved. The computational cost of training a kernel-based SVM using $N$ samples is typically between $N^2$ to $N^3$ \citep{bottou2007support}. The kernel modifies the classifier defined in equation~(\ref{classifier}) to become

\begin{equation}
\hat{y}=\sum_{i=1}^N \alpha_i y_i  K(\mathbf{x}_i,\mathbf{x}')+b.
\end{equation}

%This is of particular interest for non-linearly separated problems but comes at the cost of estimating a vector $\mathbf{\alpha}$ whose size is the number of training samples; k

\begin{figure}
  \centering
  \includegraphics[width=0.8\linewidth]{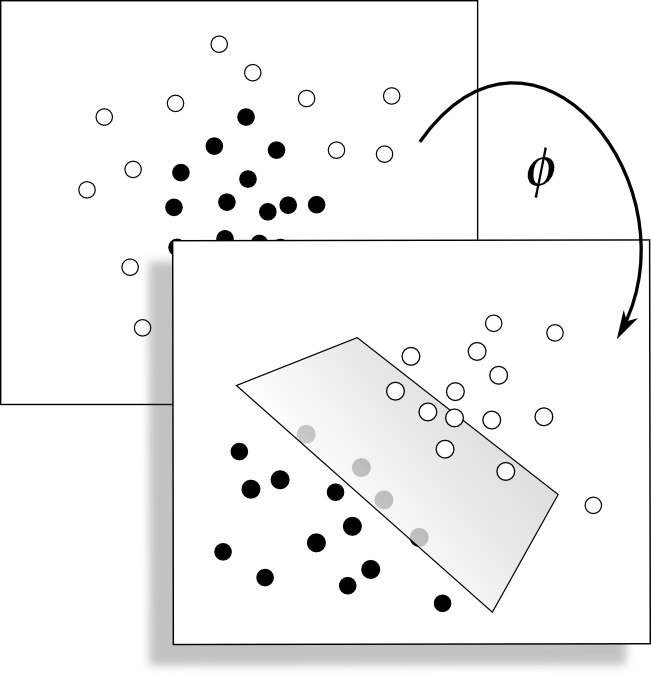}\\
   \caption{Where a linear SVM model is unable to separate data classes, a coordinate transformation $\phi$ can be used to transform each input vector into higher dimensional space, where a separating hyperplane may be found, defining a non-linear solution.}
  \label{coord}
  \end{figure}

\subsection{Mock dataset}

In order to train our SVM classifier we used simulated data products from the Bologna Lens Factory\footnote{Data products are available for download at {\tt http://metcalf1.difa.unibo.it/blf-portal/index.html}, M. Meneghetti, 2017}. The simulated images are produced using the lensing code {\sc GLAMER} \citep{2014MNRAS.445.1942M,2014MNRAS.445.1954P}, which performs ray-tracing within a halo catalogue extracted from the Millennium Simulation and identifies all regions of strong lensing.  Sources derived from real galaxy images are then added to the simulations so that they will be strongly lensed. For a detailed description of the production of the simulations see Metcalf et al. 2017, in preparation. The training data are provided as postage stamp cut-outs centred on images of either a galaxy with a lensed source or a galaxy with no lensed source. No unresolved sources were lensed in the simulated set, so lens morphology consists of arc-like objects, partial rings, or complete rings surrounding a central lensing galaxy. Late type, spiral galaxies are underrepresented in the data products, with ellipticals forming the majority of the lensing objects. An accompanying ASCII file provides classification labels for all images, along with values of the Einstein radius for all lenses. Data representing both the Euclid survey and the Kilo Degree Survey (KiDS) are available, each simulated using the corresponding PSF and survey light cones. Fig. \ref{sim_distrib} shows the Einstein radii and lensed source flux  distributions of the two training sets. The Euclid set currently contains mock images using a single, VIS, band, while the KiDS set contains the four - $g$, $r$, $i$ and $u$ - bands used in the real survey.

\begin{figure}
  \centering
  \begin{tabular}{c}
  \includegraphics[width=1\linewidth]{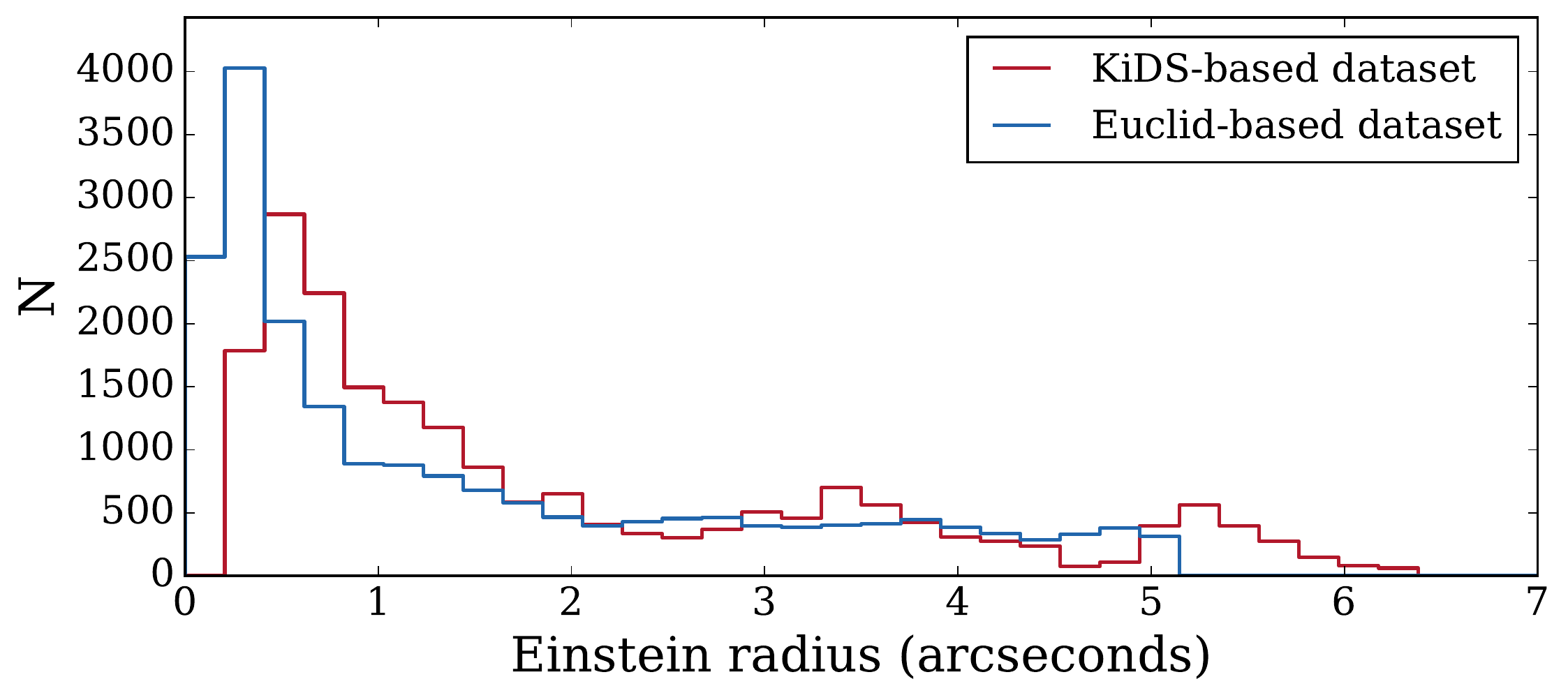}\\
  \includegraphics[width=1\linewidth]{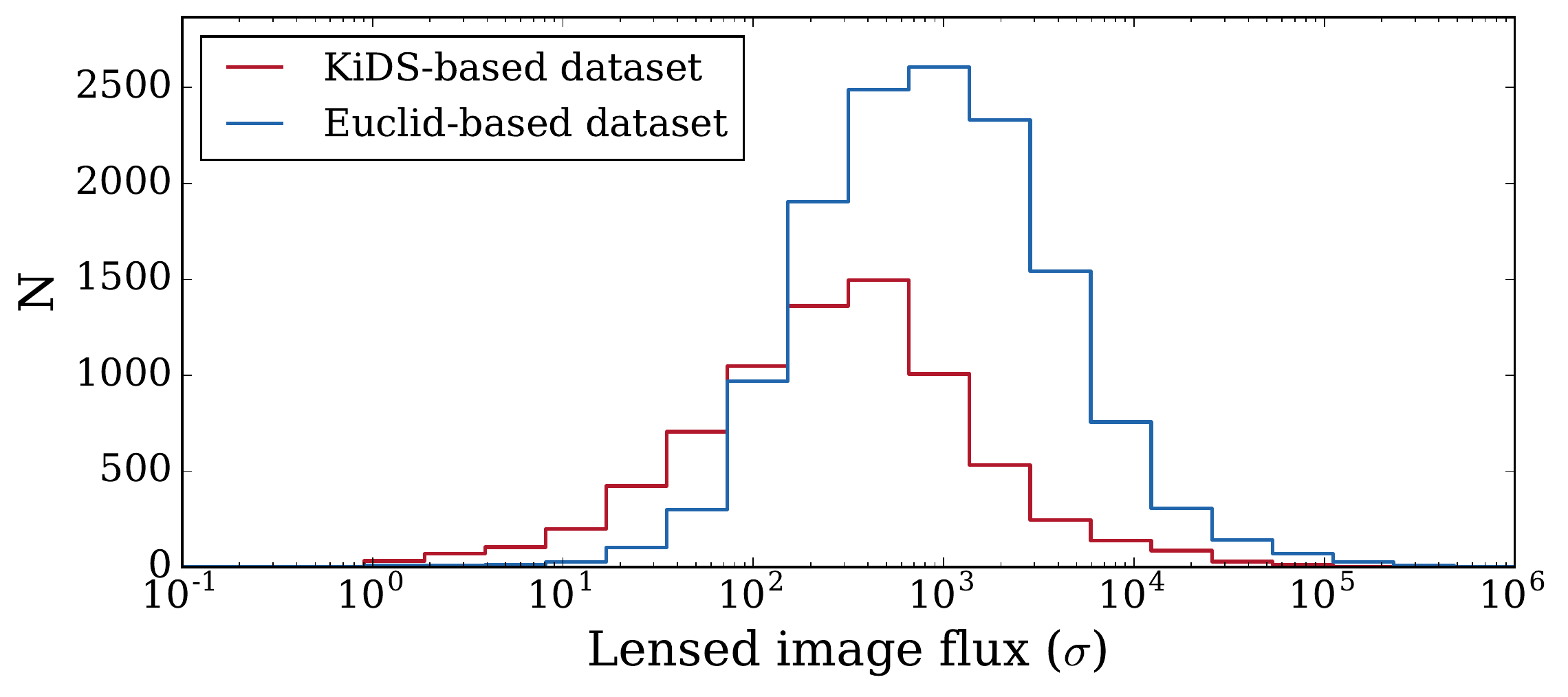}\\
  \end{tabular}
   \caption{Distributions of the Einstein radius of each lens (top) and of the flux of each lensed image (bottom), within the simulated KiDS-based (red) and Euclid-based (blue) training data provided by the Bologna Lens Factory.}
  \label{sim_distrib}
  \end{figure}

\subsection{Feature extraction}

Most image recognition machine learning methods require a two step-process. First, raw pixel values must be transformed into a set of measurable features which represent the problem under investigation. This is necessary in order to reduce variability arising  from random noise within images, ensuring that the model can generalise to new data successfully.  Once obtained, selected features are used to train the SVM for classification. Reducing raw data to a smaller set of features can also improve computational performance; for $d$ features, SVM training time increases as $d^2$ for linear SVM. Manual engineering of features can be a lengthy process, where a complex parameter space must be explored. Expert knowledge of the problem can be used to combine heuristic considerations with some automation to derive a set of informative and discriminating features. One notable exception to a manual approach is in the application of CNN, where feature learning is performed by the network itself.

In the case of galaxy-galaxy lens-finding, we are interested in discovering multiple, arc-like and ring-shaped objects in images. We also know from observation and galaxy evolution theory that a lensed source will usually appear as bluer in colour than its lensing counterpart. %despite being located at higher redshifts. 
Lensed galaxies are typically observed at an earlier stage of evolution, where specific star-formation rates are higher and young, blue, stars dominate \citep{2014MNRAS.444.2960D}. Observed lensing galaxies, on the other hand, are typically large ellipticals, where star-formation has generally ceased and stellar remnants and dust are observed as redder emissions. Selection effects may contribute to a bias in these populations: larger lensing galaxies will give rise to a greater lensing potential and a larger lensing pattern, which could be more easily discoverable in surveys than lensing by smaller blue spiral galaxies. Other anomalies include the lensing of dusty star forming spiral galaxies which can appear red, especially if observed edge-on \citep{2010MNRAS.405..783M}. We must, therefore, be careful when using colour information that we do not exclude interesting sub-populations from detection.

An ideal algorithm would not only find all lenses within a sample, but also provide reliable rejection of false positives. Sources of false positives include spiral galaxies, which contain tangentially oriented, blue structures in the form of spiral arms. In principle these can be distinguished by their multiplicity, and by their helical rather than strictly tangential shape. A further source of false positives consists of polar ring galaxies \citep{1990AJ....100.1489W,2011MNRAS.418..244M} which contain features easy to mistake for lensed background sources; a nearby example is the ring galaxy NGC~6028 \citep{1990ApJ...348..448W}.

%Our initial investigations followed the procedure used by our point-score lens-finding algorithm \rev{(see \citealt{2014A&A...566A..63J} for a comparison of this with the PCA lens finder)}, using SExtractor \citep{1996A&AS..117..393B} and GALFIT \citep{2002AJ....124..266P} to decompose each postage-stamp image into a set of objects. For each object morphological measurements were obtained, including position, radius from image centre, tangentiality and colour. Since the SVM model operates in a fixed $n$-dimensional space, features representing each image must be presented as vectors of uniform size.  To achieve this for images containing varying numbers of objects, we took statistics for each measurement made, finding the mean and variance of each value and using this set of features to train the SVM. Tests on labelled images showed some separation between lens and non-lens classes, but improvement over our point-score algorithm was only small and the loss of information from individual objects was unsatisfactory. 

In order to reduce raw images to a set of consistently relevant features, while at the same time preserving spatial localisation and colour information, we designed a bank of Gabor filters to transform all bands of each image. In 1946 \citeauthor{gabor1946theory} proposed the use of ``harmonic oscillations modulated by a `probability pulse.'" in order to reduce temporal signals to a fixed number of elementary ``quanta of information". Unlike Fourier methods, such a function could combine separate global representations of time and frequency into localised descriptions of both, enabling a new analysis of hearing sensations and the compression of speech and music. In the spatial domain, Gabor filters can correspondingly provide image processing techniques for texture classification and edge detection (e.g., \citealt{Feichtinger98a,Springer-verlag97computationalmodels}). The two-dimensional Gabor filter is composed of a Gaussian envelope modified by a complex sinusoidal plane wave. For image recognition purposes, we discard the imaginary part of the function, which corresponds only to a phase shift the frequency domain. We define the real part as:
\begin{equation}
G_c[i,j]=Be^{-\frac{(i^2+j^2)}{2\sigma^2}} \mathrm{cos}\left(\frac{2\pi}{\lambda} (i\, \mathrm{cos} \, \theta + j\, \mathrm{sin} \,\theta)\right),
\end{equation}
where harmonic wavelength $\lambda$, Gaussian spread  $\sigma$ and orientation $\theta$ define the operation performed on each point $i,j$ in an image.

\cite{Marcelja:80} showed that a family of such filters could be used to model the simple cells of the mammalian visual cortex. Further neurophysiological research  by \cite{daugman1985uncertainty} found that the parameters $\lambda$ and $\sigma$ are closely correlated and that, for simple visual cells, the spatial frequency bandwidth determined by the ratio $\sigma/\lambda$ occupies only a limited range. In constructing our filter bank, we chose to apply a similar bandwidth restriction to ensure consistent localisation in the frequency domain.

\begin{figure}
  \centering
      \includegraphics[width=9cm]{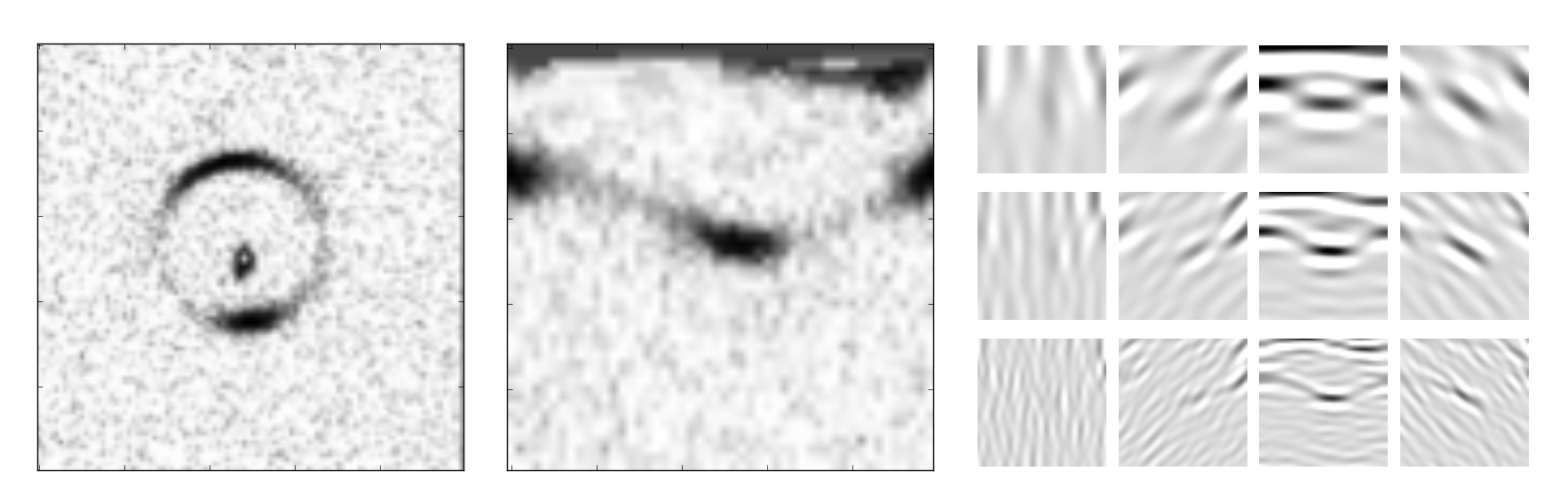} 
  \caption{ Response of a set of Gabor filters (right) after convolution with a polar transformed image (middle) of an Einstein ring (left). The strongest filter response is seen in the orientation perpendicular to the radial direction and at the frequency most closely matching that of the ring.}
 \label{polarfilter}

\end{figure}

Before convolution with the filter set (see Fig.~\ref{polarfilter}), we first applied a polar transform to each image, aiming to exploit the edge-detection ability of the Gabor function in order to pick out tangential components typical of galaxy-galaxy lensing. With each image centred on a galaxy, the origin is located at the centre of the image. The image is then transformed by converting the two-dimensional Cartesian coordinates, x and y, of individual pixels into polar coordinates, $\theta$ and $\rho$. $\theta$ represents the counter-clockwise angle in the x-y plane measured in radians from the positive x-axis. $\rho$ is the radius from the origin to a point in the x-y plane and is limited to a value of half the image width.

The generality of spatial information obtained from the filter set allows the SVM model enough freedom to detect which features define the problem best. The responses of the applied filters are measured by calculating moments for each filtered image. Specifically, for the image pixel values $x_1,...,x_N$, we evaluate the following statistical moments: mean, variance, skew, kurtosis and local energy (see Table~\ref{statstab}). These statistics form our final features on which the SVM would train and classify. We perform standardisation on each feature, setting the mean to zero and the variance to one. Without this scaling, features of higher variance may dominate the final SVM model.

\begin{table}

	\centering

	\begin{tabular}{lccr} % four columns, alignment for each
		\hline
		Mean  & \( \mu_1(x_1,...,x_N)= \frac{1}{N}\sum_{j=1}^{N}x_j \)\\[1.5em]
		Variance & \( \mu_2(x_1,...,x_N) = \frac{1}{N-1}\sum_{j=1}^{N}(x_j-\mu_1)^2
 \) \\[1em]
		Skew & \( \mu_3(x_1,...,x_N)= \frac{1}{N}\sum_{j=1}^{N}\left[\frac{x_j-\mu_1}{\mu_2}\right]^3 \)\\[1em]
		Kurtosis & \( \mu_4(x_1,...,x_N) = \left\lbrace  \frac{1}{N}\sum_{j=1}^{N}\left[\frac{x_j-\mu_1}{\mu_2}\right]^4 \right\rbrace \) \\[1.5em]
        Local energy & \( E_s(x_1,...,x_N)= \sum_{j=1}^{N}x_j^2 \) \\
        
		\hline
	\end{tabular}
        
    	\caption{The response of each Gabor filter was measured by calculating the above  statistics for each filtered image.}
        \label{statstab}
\end{table}

%mean
%\begin{equation}
%\mu_1(x_1,...,x_N)= \frac{1}{N}\sum_{j=1}^{N}x_j,
%\end{equation}
%variance
%\begin{equation}
%\mu_2(x_1,...,x_N) = \frac{1}{N-1}\sum_{j=1}^{N}(x_j-\mu_1)^2,
%\end{equation}
%skew
%\begin{equation}
%\mu_3(x_1,...,x_N)= \frac{1}{N}\sum_{j=1}^{N}\left[\frac{x_j-\mu_1}{\mu_2}\right]^3,
%\end{equation}
%and kurtosis
%\begin{equation}
%\mu_4(x_1,...,x_N) = \left\lbrace  \frac{1}{N}\sum_{j=1}^{N}\left[\frac{x_j-\mu_1}{\mu_2}\right]^4 \right\rbrace.
%%\end{equation}
%in addition to the local energy, defined as the sum of squared pixel values:
%\begin{equation}
%\mathrm{local}\, \mathrm{energy}(x_1,...,x_N)= \sum_{j=1}^{N}x_j^2.
%\end{equation}
%\rf{RF:
%Finally scatterplot of the two most discriminant features could (with the moments above) would also be nice. Or else a PCA projection of TSNE projection of the data to see the neighborods of the classes. I can do that if you send me the training matrices it can also help you understand different class clusters . All the tools are provided in sklearn}

%\begin{figure}
%  \centering
    
%      \includegraphics[width=1\columnwidth]{gray2.png} 
%  \caption{ }
% \label{gabor2}

%\end{figure}

\subsection{Feature selection and dimensionality reduction}

The performance of a supervised machine learning model can be assessed by evaluating the accuracy of the model on an unseen test set of data. Using a unique set of data for testing ensures that the model is learning features that are relevant to the classification problem, rather than simply the features of the training data itself. We can characterise the difference between training accuracy and test accuracy and use this to understand how well the model will generalise to new data. Structural risk minimisation \citep{Vapnik:1995:NSL:211359} evaluates the bound
\begin{equation}
\mathrm{test\, error = train \,error} + f(N,h,p),
\label{bound}
\end{equation}

where $N$ is the number of samples in a training dataset and $h$ is the measure of model complexity.  Probability of this bound failing is included as $p$. Increasing the number of training samples reduces this bound; increasing model complexity results in a trade-off between training error and test error. The Vapnik-Chervonenkis (VC) dimension provides measure of model complexity. It is defined as the number of data-points that a model can learn perfectly - or 'shatter' - for all possible assignments of labels. A model that can shatter its dataset is too complex; if it can learn any random association of data-points then generalisation to unseen data will be poor \citep{Burges1998}. 
 
For a linear SVM, a hyperplane model results in a VC dimension of $k+1$ for $k$ dimensions; adding more features increases model complexity and transformation into non-linear space will further increase complexity. However, the  maximum margin approach used by the SVM ensures that the simplest model possible is used. Therefore, we are free to use large numbers of features, provided we have a correspondingly large training set and that proper regularisation is implemented. We used several methods to evaluate feature importance and avoid overcomplexity.

\begin{figure}
  \centering
      \includegraphics[width=1\linewidth]{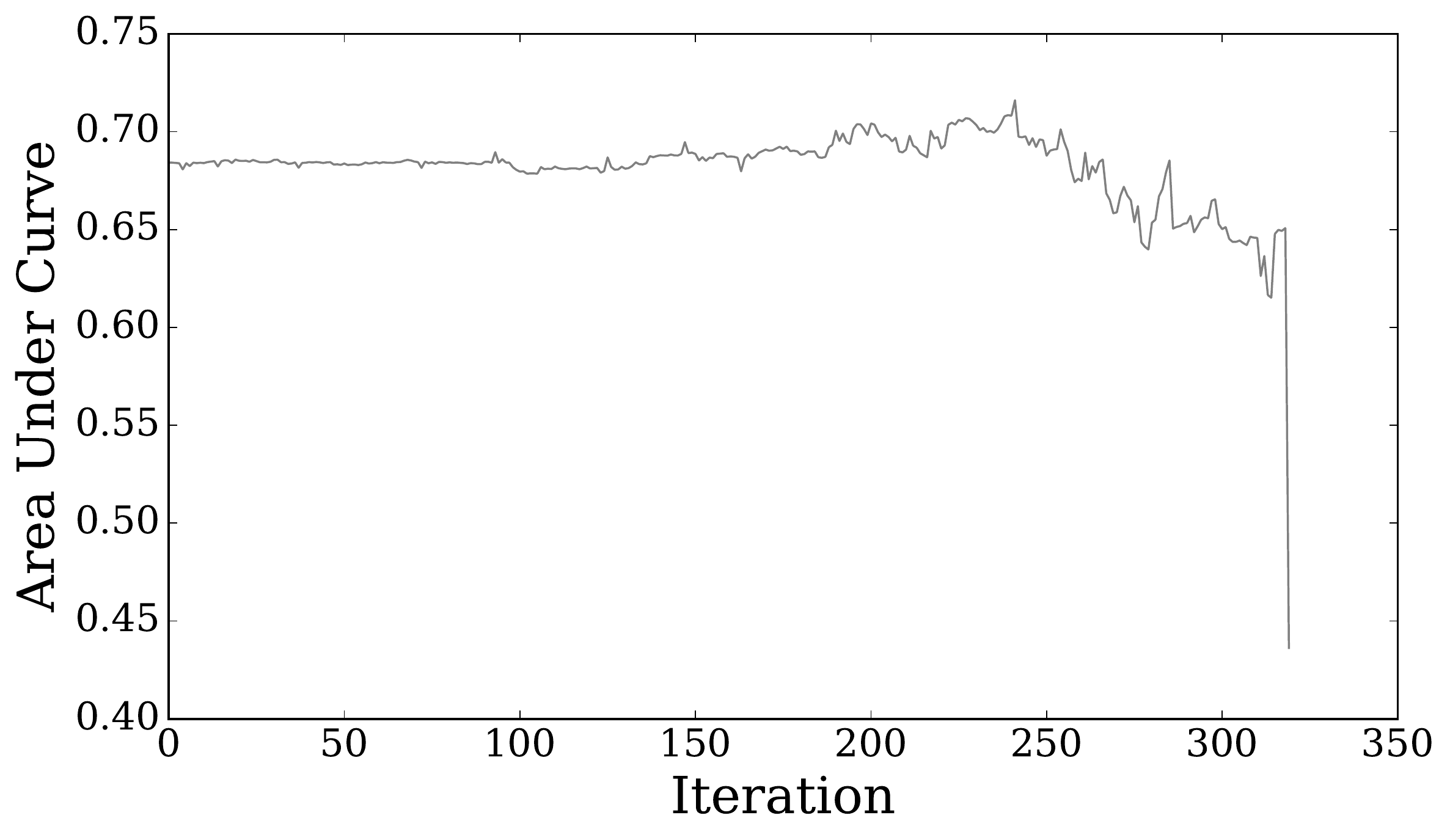} 
  \caption{Recursive elimination of features. Feature importance is evaluated using weights assigned by a linear SVM. In each iteration the feature with the lowest importance is removed. Some improvement in classification performance is seen but the method can be unstable. Feature reduction may not be necessary with a properly regularised SVM kernel and a large enough training set. } 
 \label{recursive}
 
 \end{figure}
 
\subsubsection{Feature selection methods}

Recursive feature selection is a brute-force technique which uses weights assigned by a linear SVM to determine the importance of features. Removing features of the lowest importance and iterating over this procedure can optimise results. We found some improvement with this approach. However, the method can be unstable; initial weights will, by definition of the problem, be under-optimised, so feature importance  will be unreliable. The iterative procedure also risks tuning the model to the specific dataset with which features were selected, and non-linear relationships are ignored. Results obtained from this approach should be used with caution.

Principal component analysis (PCA) determines covariant eigenvectors within a dataset and re-projects the whole dataset in order of eigenvalues.  Re-projecting SVM features along directions of high covariance can remove redundant and duplicated features. Re-projecting onto 2D or 3D space allows for visualisation of the feature space. However, PCA considers only linear relationships, ignoring the possibly nonlinear nature of data structure. While we found inspection of the highest principle components useful in evaluating feature importance, we found no improvement in accuracy when using PCA to reduce the feature set; retaining non-linearities that can be captured by an SVM is essential.  Kernel PCA can apply a chosen kernel to the data before evaluating the projection of transformed data onto principal components \citep{Scholkopf99kernelprincipal}. The selection of features could then be automated by using Multiple Kernel Learning which can learn the best combination of PCA kernels from a predetermined set \citep{Gonen}. This is rather computationally costly but is a potential avenue for future work. 

t-distributed stochastic neighbour embedding (t-SNE) can be used as an alternative to PCA as a data-visualisation tool \citep{ictdbid:2777}. t-SNE goes beyond straightforward projection, embedding  features into lower dimensional space while preserving non-linear relationships by applying different transformations to different regions. The mapping is made by measuring similarities between points in space, and using gradient descent to minimise the Kullback-Leibler divergence \citep{kullback1951,Kullback68informationtheory} between joint probabilities of the high-dimensional points and joint probabilities of the embedded points. As a probabilistic method, t-SNE does not permit the mapping of new data onto the lower dimensions containing the embedded data. Therefore, the output from t-SNE cannot be used directly, but can be used as a visual guide to intuit the selection of features. Fig.~\ref{tsne} shows a visualisation of the features extracted from the  KiDS-based mock data set. A reasonable division between classes is seen.

\begin{figure}
  \centering
      \includegraphics[width=8.5cm]{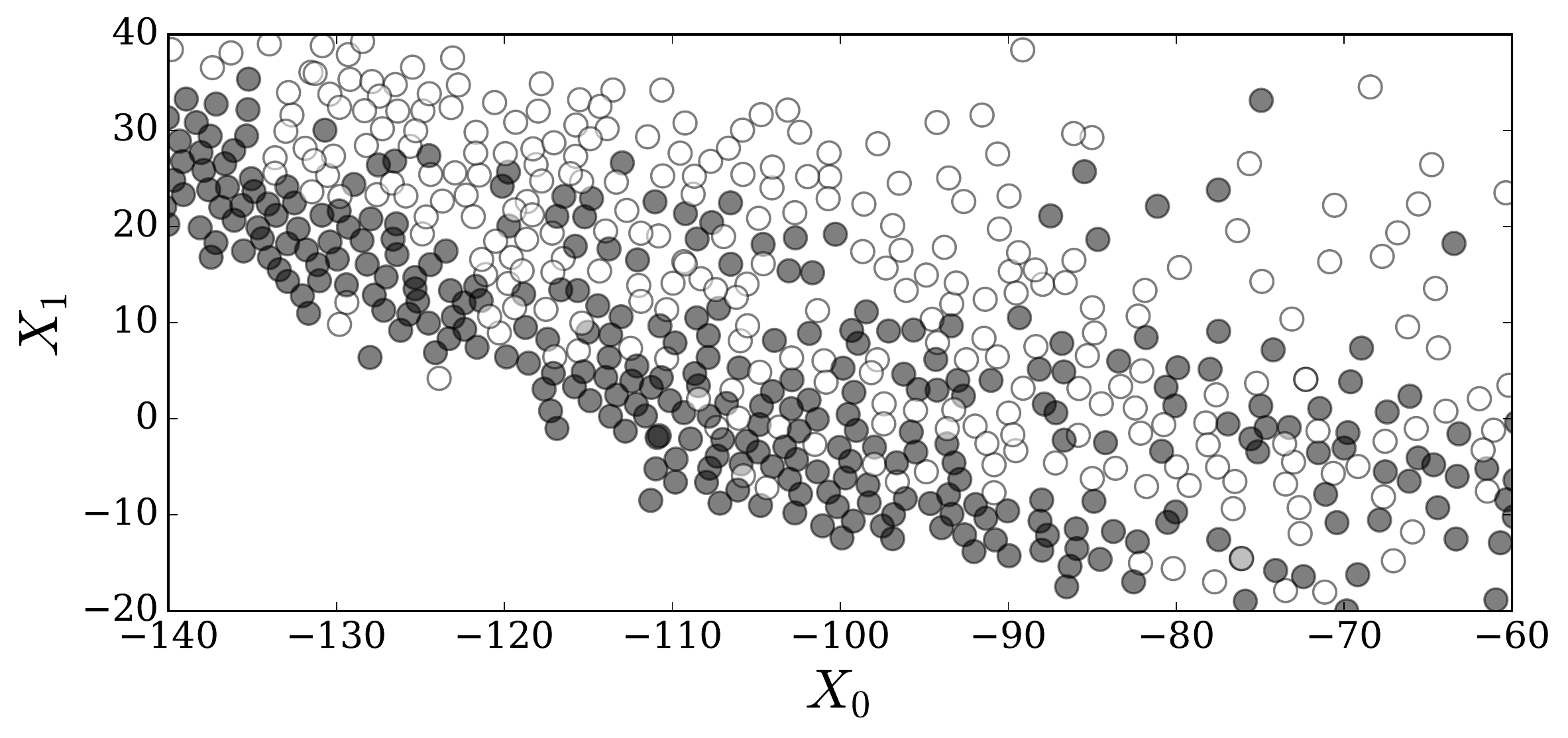} 
  \caption{A visualisation of the features extracted from the KiDS-based mock dataset. Individual data samples containing 1600 features have been embedded into 2-dimensional space using t-distributed stochastic neighbor embedding (t-SNE). The values $X_0$ and $X_1$ represent combinations of features obtained by measuring similarities between points in space, and using gradient descent to minimise the Kullback-Leibler divergence \citep{kullback1951,Kullback68informationtheory} between joint probabilities of the high-dimensional points and joint probabilities of the embedded points. The mapping preserves non-linearities and displays some separation between the class of images which contain lensed sources (open circles) and the class without lensed sources (filled circles). } 
 \label{tsne}

\end{figure}

Our final feature selection approach involved sub-sampling the feature set before applying selection algorithms. Such stability selection methods evaluate feature importance within randomised subsets before calculating an aggregate score for each feature. In our investigations we used the randomised lasso method developed by \cite{2008arXiv0809.2932M}. Lasso - least absolute shrinkage and selection operator - is a regression analysis technique developed to improve the results of least squares methods by including a regularisation parameter which can penalise overcomplexity. By assigning regression coefficients to each feature and insisting that the sum of absolute coefficients remains less than a fixed value, lasso will force many coefficients to zero, resulting in a sparse model with only a small number of features achieving a non-zero coefficient  \citep{Tibshirani94regressionshrinkage}. Lasso performs well when applied to features of low mutual coherence, but begins to fail in the case of correlated features, picking one at random. Randomised lasso tackles this problem  by applying lasso to random subsets of features, iterating over the whole feature set many times, counting the number of times a feature is selected. While the method is again confined to the linear regime and cannot detect important non-linear relationships, we found it useful to gain general insight into the importance of different families of features. 
%.(and \cite{bach:hal-00354771}) bootstrap lasso\citeauthor{2011arXiv1104.3398W} other random lasso paper

Using the randomised lasso method, we evaluated feature importance as a function of Gabor kernel frequency and as a function of Gabor kernel orientation, as well as for each moment derived from filter responses. We found that classifier performance favoured Gabor frequencies of the order 0.1 pixel$^{-1}$. We used a log series of frequencies around this value in order to maximise signal response.   Predictably, rotation values which were roughly parallel to the theta direction of the polar transformed image were favoured. However, other directions scored not significantly lower, highlighting the ability of the Gabor feature set to generalise well to other morphologies. We found that the higher order moments, skew and kurtosis, derived from filter responses were found to be of greater importance than the mean and standard deviation.

\begin{figure}
  \centering
      \includegraphics[trim={0.5cm 1.5cm 1cm 1.5cm},clip,width=8.5cm]{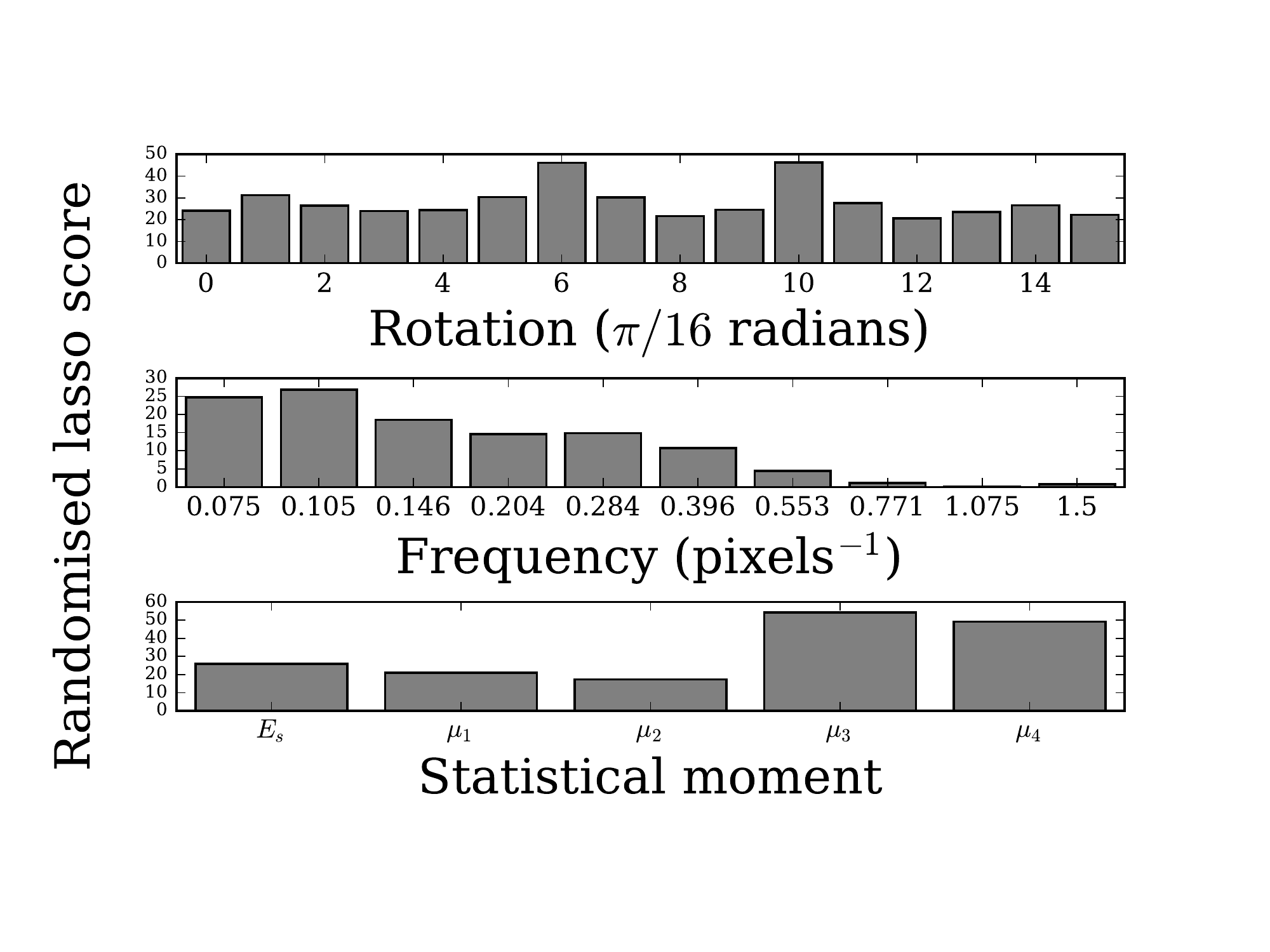} 
  \caption{Feature importance measured using a randomised Lasso regression to sub-sample the feature space. Scores reflect the number of times each feature was selected by a sparse model. Arranged into families of Gabor filters, an interesting dependence on rotation is seen, where feature importance peaks for angles not quite parallel to the tangential direction of the polar-transformed image. A frequency dependence is also seen, and importance extends further into the lower frequency domain than the computer used for development would permit in a reasonable time interval; as frequency decreases, the size of the convolution kernel increases. This parameter space can be explored using a faster machine, writing in lower-level languages, or exploiting GPU computing. Higher-order statistics are noted to be of significant importance.  } 
 \label{family}

\end{figure}

After investigations using a combination of the above techniques, we determined a final set of Gabor filters as summarised in Table~\ref{filtertable}. The importance of different feature families is illustrated in Fig.~\ref{family}. We note that the  feature space would benefit from extension into the lower-frequency Gabor filter range. Since a lower frequency results in a larger convolution kernel we were ultimately limited by the processing speed of the modest equipment used, but the use of more powerful machines would make this feasible.

\begin{table}
	\centering

	\begin{tabular}{lccr} % four columns, alignment for each
		\hline
		 & rotation (radians)&frequencies (pixel$^{-1}$)\\
        \hline
		Space &\(\sum_{n=0}^{15} \frac{\pi}{16}n\) & \( \sum_{n=0}^{9} 10^{\,\log_{\scalebox{0.5}{10}}a +(\log_{{\scalebox{0.5}{10}}} b \,-\, \log_{\scalebox{0.5}{10}} a/9)\,n} \) \\[1em]
		Ground  &\(\sum_{n=0}^{7} \frac{\pi}{16}n \) & \( \sum_{n=0}^{9} 10^{\,\log_{{\scalebox{0.5}{10}}} a +(\log_{\scalebox{0.5}{10}} b\, -\, \log_{\scalebox{0.5}{10}}a/9)\,n} \)  \\

		\hline
	\end{tabular}
    \caption{Gabor filter sets are described. The filters used are defined by both the frequency of the sinusoidal component and the  rotation of the kernel in the 2D plane. A log distribution of frequencies was used in order to capture more information from the low frequency domain.  Frequencies ranged from $a=0.075$ to $b = 1.5$. Rotations were spaced linearly between 0 and $\pi$ radians. Filters were applied to the single band images of the space-based data and to all four bands of the ground-based data.}
    \label{filtertable}
\end{table}

\subsection{Kernel selection}

We have chosen a feature set with the aim of balancing the ability of the model to pick out important relationships in the training data, while still generalising well on unseen data. We can further guard against poor generalisation by using a large training set. This problem can be thought of as one of bias versus variance. A low - or misrepresented - number of features could result in bias within the classifier. The model may be underfit, ignoring relevant relationships between features and poorly representing the problem. On the other hand, using insufficient samples to train a classifier which is too complex will result in a high degree of variance; the model will suffer from overfitting if it is too sensitive to small fluctuations in the training data. SVM have an additional protection against this bias-variance problem. By carefully tuning a small group of regularisation hyperparameters, the SVM can reduce the bound on the risk (equation~(\ref{bound})) and remain robust to overfitting even when a very large feature set is used.

%\rev{$\gamma$ and $C$ (Fig.~\ref{paramselect})}, 

\begin{figure}
  \centering
  \begin{tabular}{c}
  \includegraphics[width=1\linewidth]{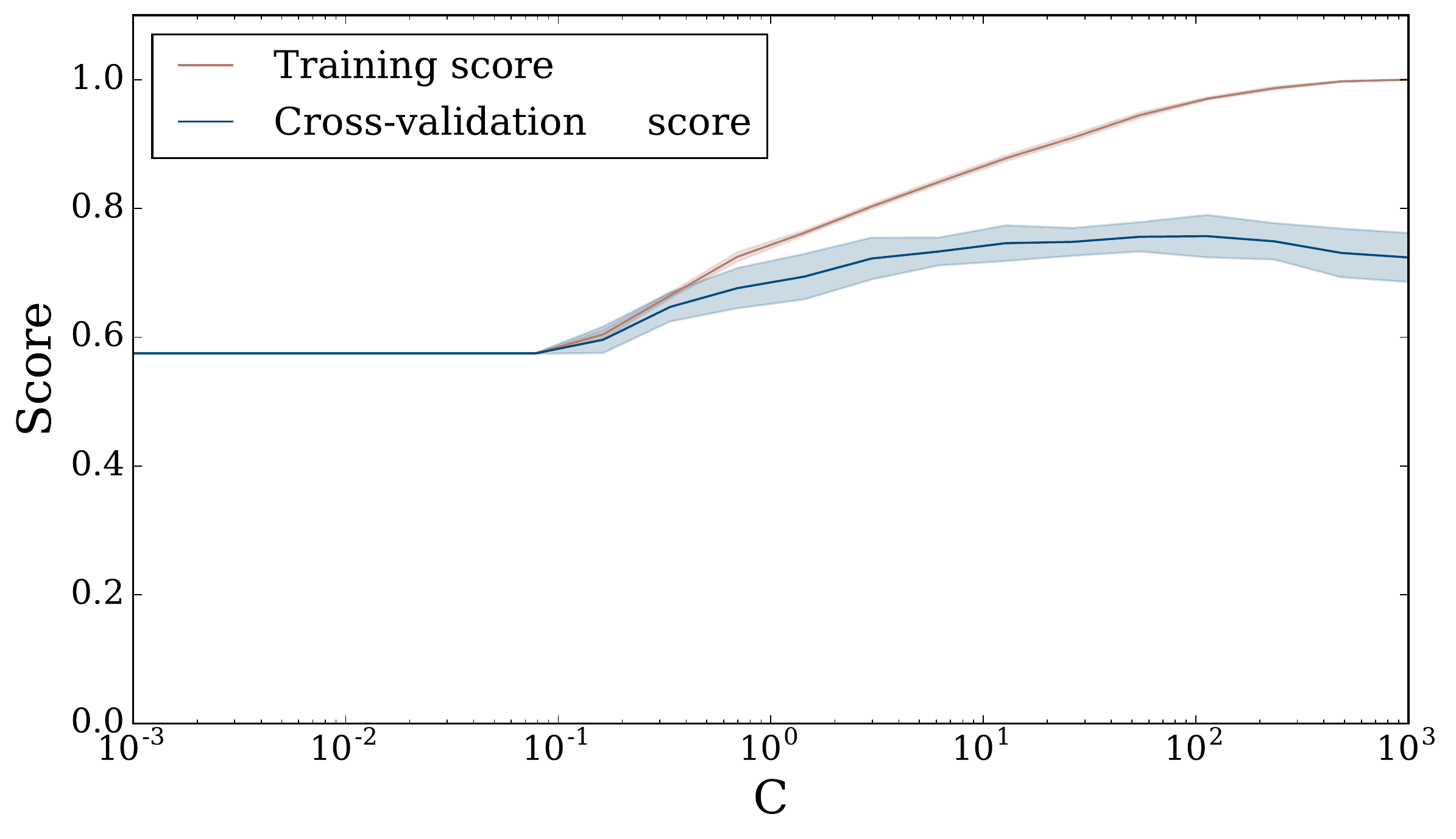}\\
  \includegraphics[width=1\linewidth]{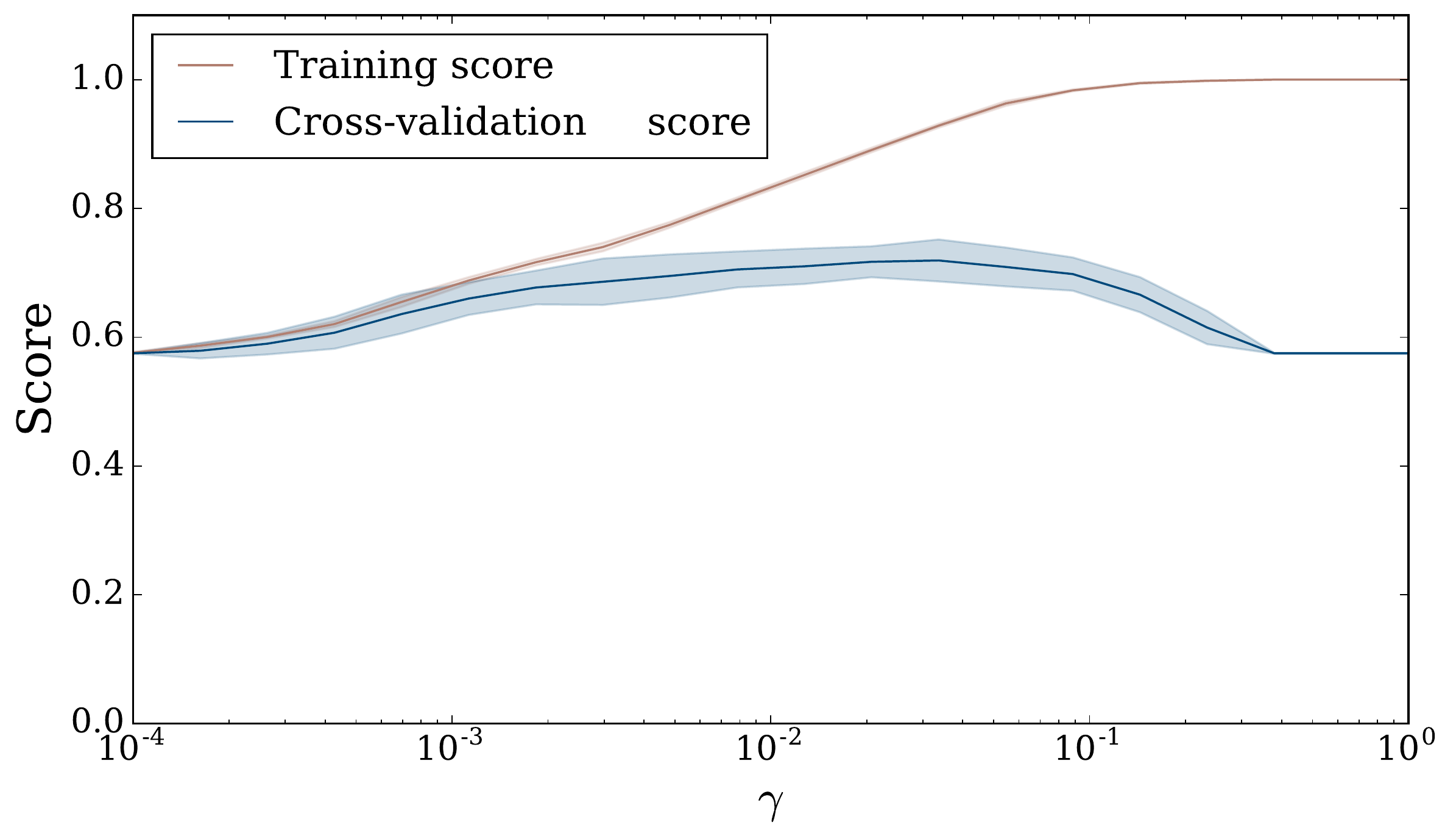}\\
  \end{tabular}
  \caption{ Validation curves produced by varying the SVM kernel parameters $C$ (top) and $\gamma$ (bottom). Classification performance of the finder is tested on the training data set (red curves) and on a held-out, validation, set (blue curves). Low values of both parameters result in underfitting of the model. With increasing values the model becomes more complex and scores for the training set to begin diverge from those for the test set. Beyond optimal values overfitting occurs and performance on the test set begins to fall.}
  \label{paramselect}
  \end{figure}

%\begin{figure}
%  \centering
%      \includegraphics[width=1\columnwidth]{gridsearch.pdf} 
%  \caption{}
% \label{gamma}

%\end{figure}
Selecting an appropriate kernel for an SVM is an important first step. In our implementation, a linear kernel was unable to separate the data classes and non-linear kernels were investigated.  After experimentation, we chose a radial basis function (RBF) kernel. Introduced by  \citeauthor{Smola:1998:CRO:294223.294228} while investigating the correspondence between support vectors kernels and regularisation operators, the Gaussian RBF kernel is defined as
\begin{equation}
K(\mathbf{x}_i,\mathbf{x}_j) = \mathrm{exp}\left(-\gamma\|\mathbf{x}_i-\mathbf{x}_j\|^2\right),
\end{equation}
where $\gamma = 1/2\sigma^2$. The RBF kernel can map to an infinite number of dimensions. We can use $\gamma$ as a regularisation hyperparameter to control the influence of a single training sample on the rest of the set. Overfitting can be avoided by reducing $\gamma$ so that individual support vectors are influenced by those within a greater radius (see Fig.~\ref{paramselect}). Another hyperparameter, C, controls the trade-off between misclassification of samples and the simplicity of the separating hyperplane by constraining the weighting coefficient $\alpha_i$ of each training sample to $0\geq \alpha_i \geq C$; a large C will tend towards the selection of a greater number of support vectors, resulting in a more complex model and the risk of overfitting. 

Brute-force optimisation over the parameter space provides an easy, if time-consuming, way to select the best kernel hyperparameters. We used a grid search method to find the best ranges for values of $\gamma$ and $C$.  Subsequent searches operate over the optimal regions to increase granularity. The data classes remained inseparable under the final model, but performance was improved over that of the linear kernel.

%\begin{figure}
%  \centering
%      \includegraphics[width=1\columnwidth]{einscore.pdf} 
%  \caption{}
% \label{einsteinvsscore}

%\end{figure}

%In statistics and machine learning, the bias–variance tradeoff (or dilemma) is the problem of simultaneously minimizing two sources of error that prevent supervised learning algorithms from generalizing beyond their training set:

%    The bias is error from erroneous assumptions in the learning algorithm. High bias can cause an algorithm to miss the relevant relations between features and target outputs (underfitting).
%    The variance is error from sensitivity to small fluctuations in the training set. High variance can cause overfitting: modeling the random noise in the training data, rather than the intended outputs.

%these points explain why having too many features leads to overfitting - i.e. the variance is higher

%can invoke the bias-variance tradeoff to explain the effectiveness of heuristics in human learning.

\subsection{Results}

It is important to define some sensible metrics so that the success of the model can be evaluated for the specific requirement of the classification problem. A fully separable dataset is the goal, and achieving this would indicate that the problem has been solved. In practice, however, classifiers are often unable to fully separate the data, and  success can be defined in various ways. Of particular interest is the trade-off between the purity of the results - that is, how many of the positive classifications are correct - and the the completeness of the results - or how many of the positive samples are correctly identified. In the case of lens finding, there are  science cases to be made for either leaning: statistical techniques may require very large samples of lens systems at the cost of many false positives, but science involving the observations and modelling of individual lens systems can demand a high level of purity. In practice, high purity is needed within very large samples, otherwise candidate follow-up becomes a prohibitively expensive process. It becomes necessary to preserve this scope so that evaluation with respect to various applications can be considered. %\rf{RF: Note that this machine learning problem (minimising one type of error with constraint on another one) à is called Neyman Pearson classification and several ML technique exists to address it with statistical guaranties. Still SVM seems to works well here. }

In the non-separable case, the SVM is able to provide classification scores for an unseen dataset using the distance of each sample from the separating hyperplane. A larger absolute score reflects a greater confidence in the classification in the sample. Plotting these scores, we can quickly see how well two classes of data have been separated but values of purity and completeness will vary according to which score threshold we choose (see Fig.~\ref{cvscores}). 
Using the labelled test data and defining true positive rate (TPR) and false positive rate (FPR) as
\begin{equation}
\mathrm{TPR} = \frac{\sum \mathrm{true\,positives}}{\sum \mathrm{true\,positives}+ \sum \mathrm{false\,negatives} }
\end{equation}
and
\begin{equation}
\mathrm{FPR} = \frac{\sum \mathrm{false\,positives}}{\sum \mathrm{false\,positives}+ \sum \mathrm{true\,negatives} },
\end{equation}
a Receiver Operating Characteristic (ROC) curve is created by plotting TPR against FPR for increasing thresholds. The area under the curve (AUC) provides a single measure of general performance of the classifier.  

We used 20000 training samples each to train classifiers to find lenses in KiDS observations and in Euclid observations. We performed cross validation, maximising the use of the training data by iterating over $n$ folds; training was performed on $n-1$ folds and results tested on 1 held-out fold during each iteration. Final scores achieved were an AUC of 0.89 for the single-band Euclid dataset and 0.95  for the four-band KiDS dataset (see Fig.~\ref{roc}). From this difference and from tests using only single bands of the KiDS dataset we conclude that colour information is important for successful classification.

\begin{figure}
  \centering
  \begin{tabular}{c}
  \includegraphics[width=1\linewidth]{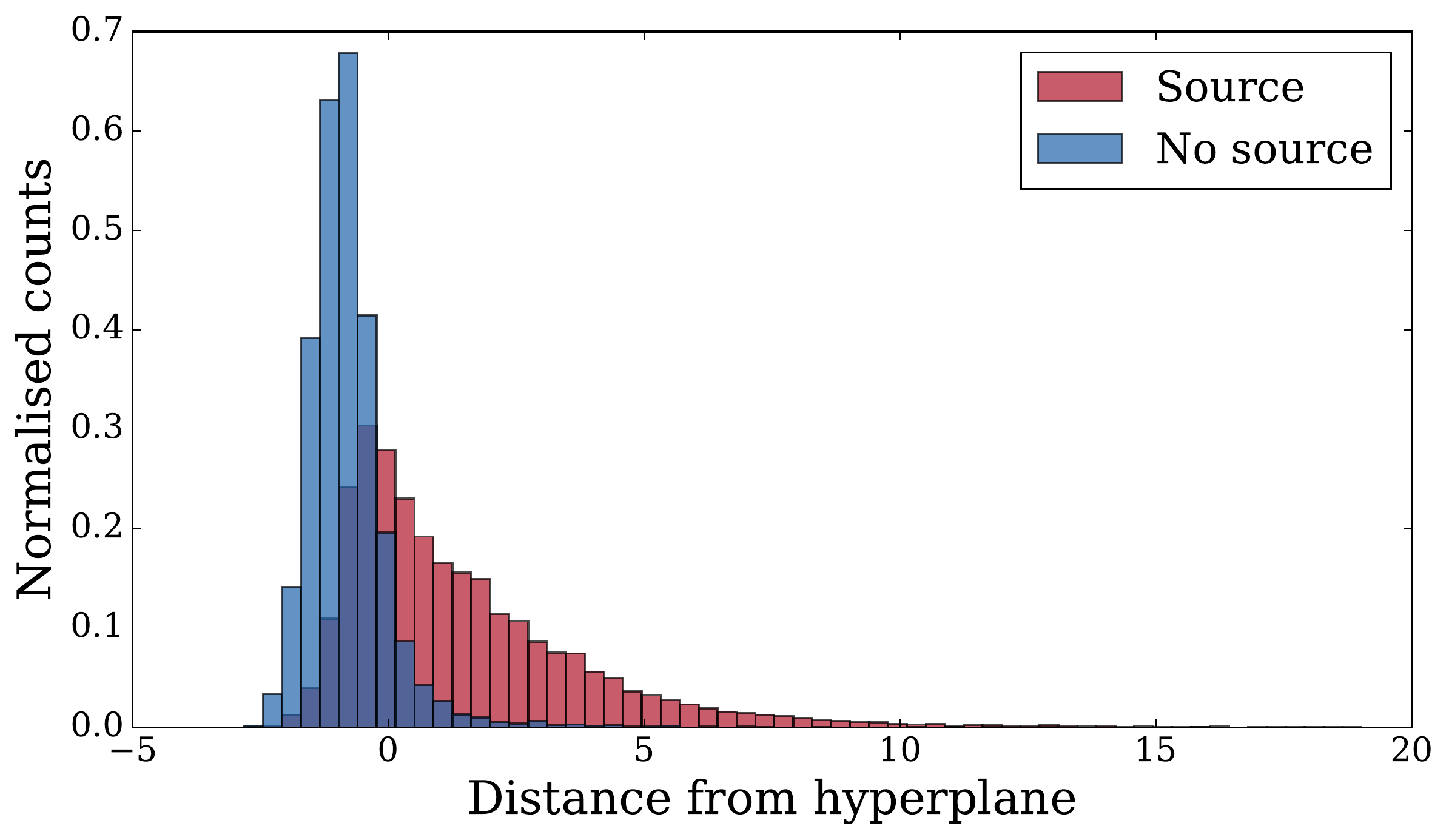}\\
  \includegraphics[width=1\linewidth]{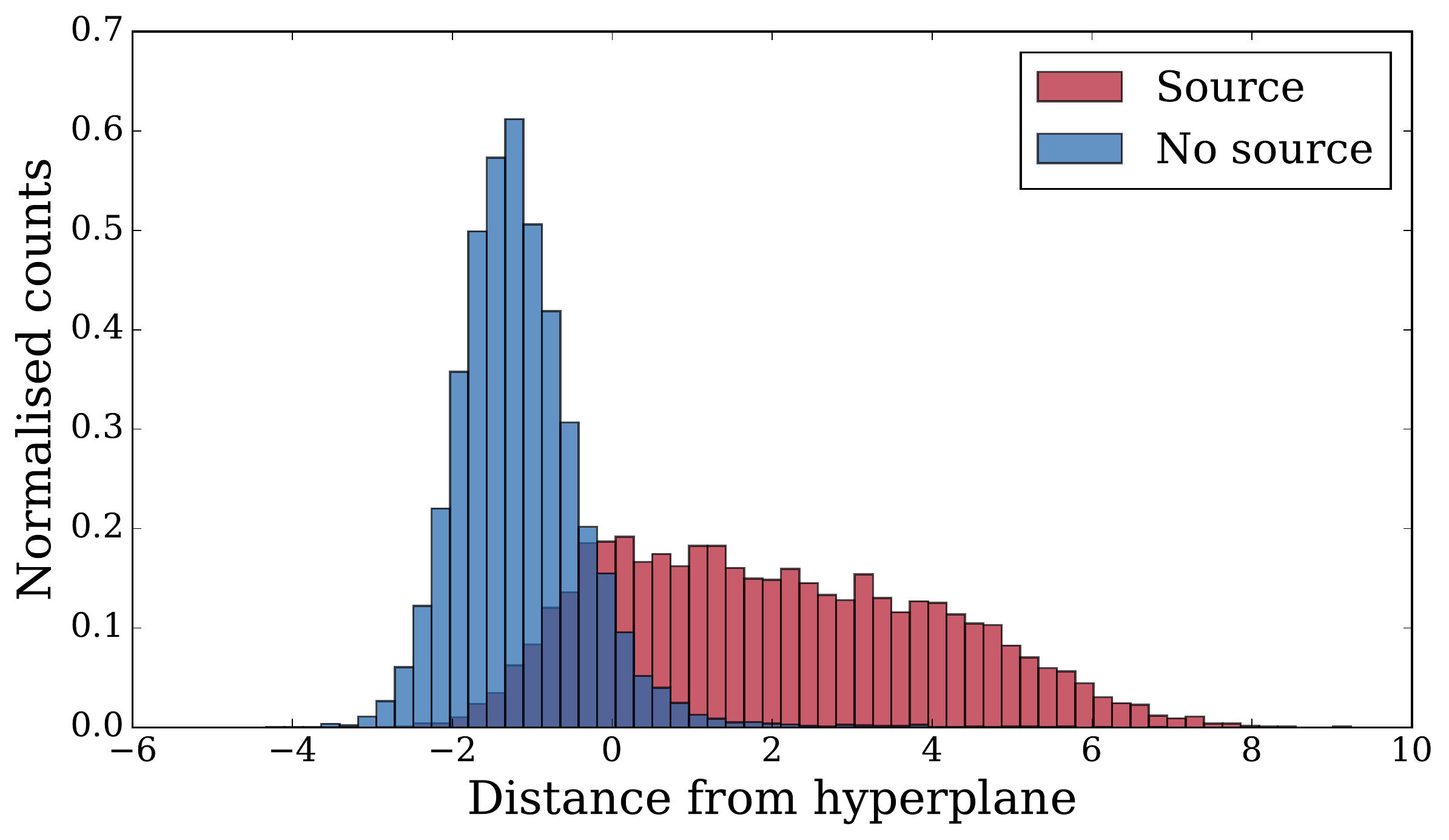}\\
  \end{tabular}
  \caption{Classification scores from application of the Gabor-SVM method to unseen mock Euclid data (top) and  unseen mock KiDS data (bottom). The SVM represents individual data samples as vectors. A hyperplane separating two classes is defined using training data to maximise the margin between ``support vectors'' on the inner edge  of each class. Scores represent the distance of each individual unseen sample from the hyperplane. }
  \label{cvscores}
  \end{figure}

\begin{figure}
  \centering
  \begin{tabular}{c}
  \includegraphics[width=1\linewidth]{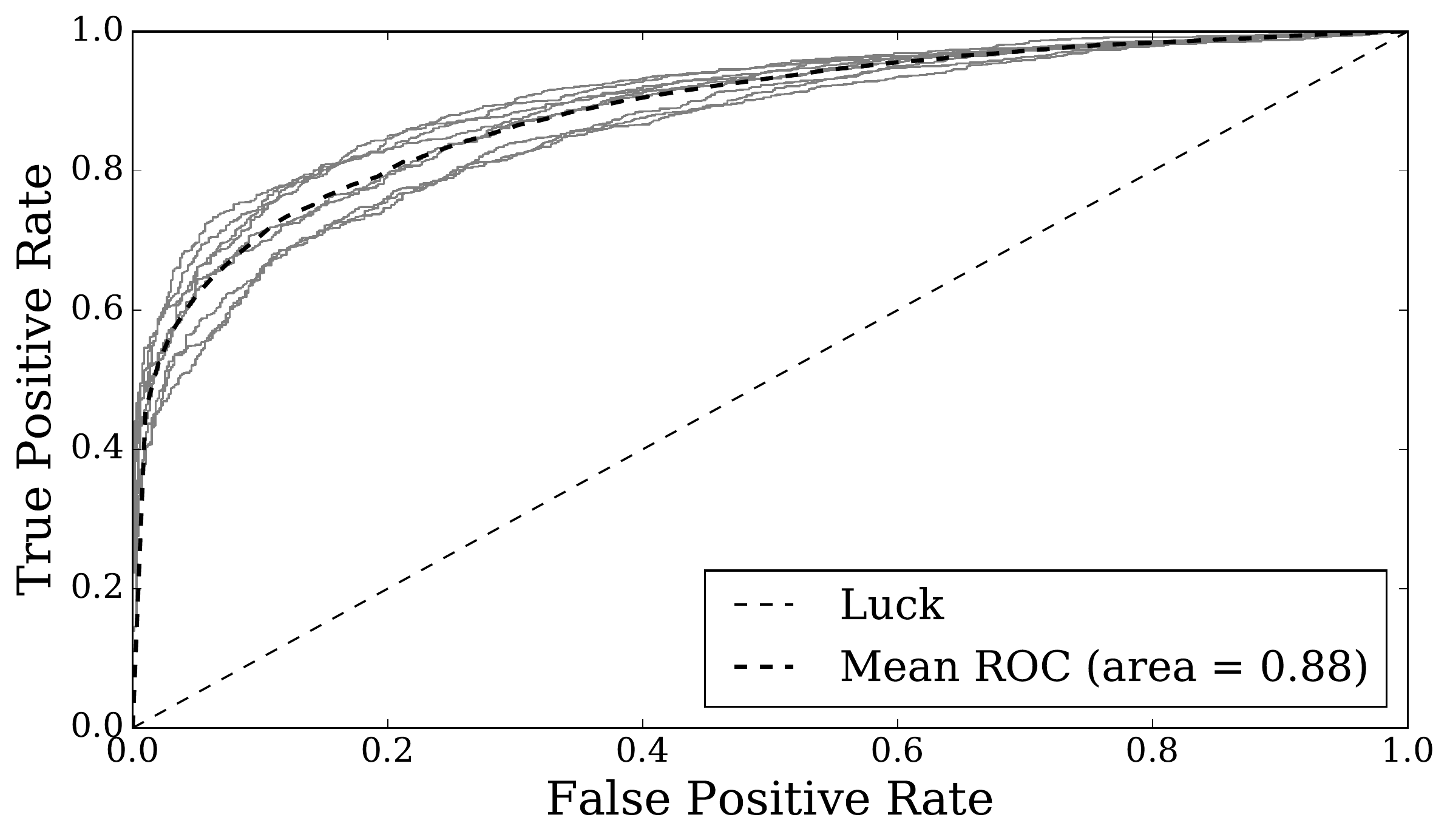}\\
  \includegraphics[width=1\linewidth]{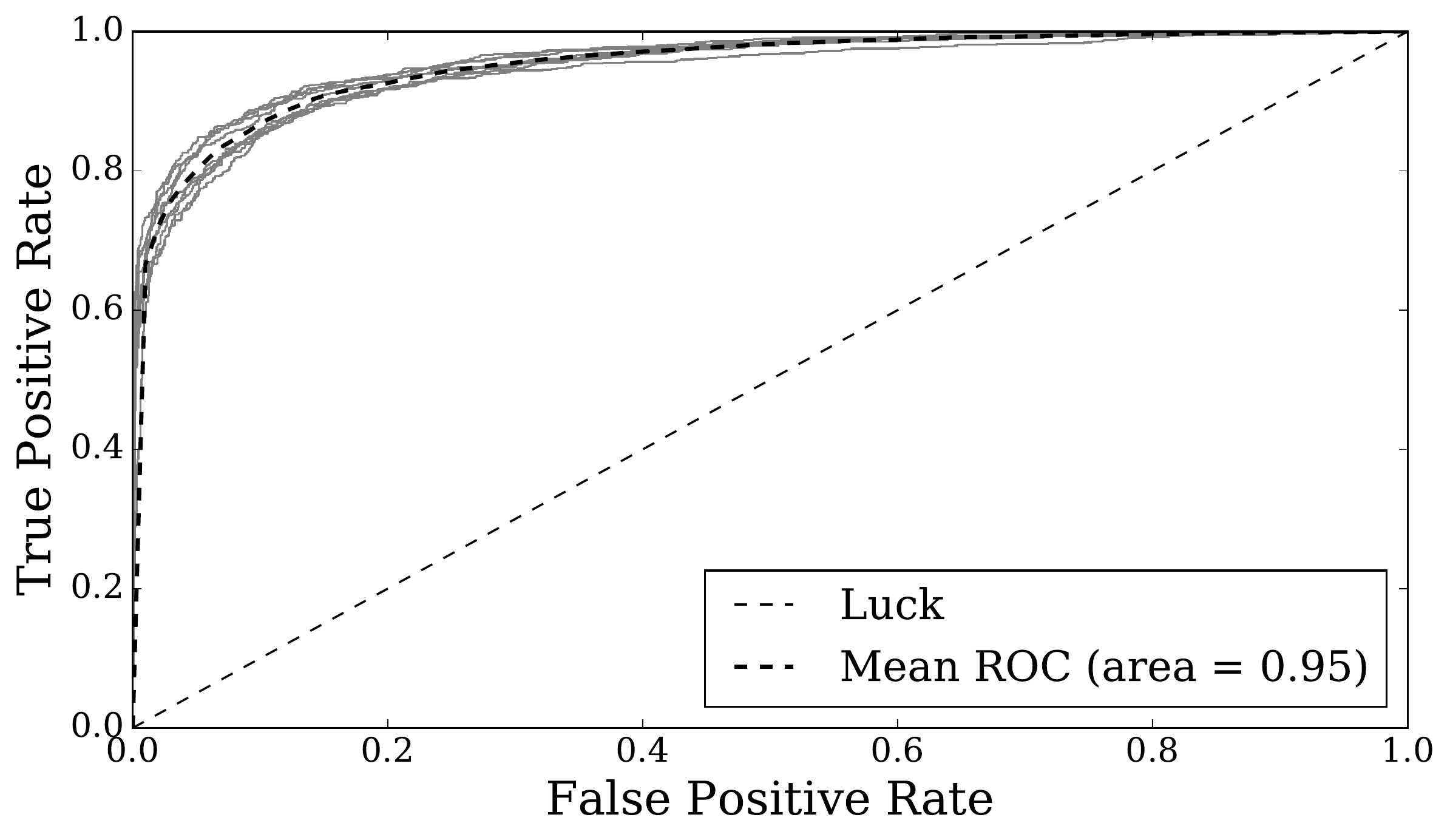}\\
  \end{tabular}
  \caption{Receiver operation characteristic (ROC) curves illustrate the trade-off between the ``purity'' ($1-$ False Positive Rate) of a classifier and its ``completeness'' (True Positive Rate). The area under the curve represents the overall perfomance and a perfect classifier will have an area of 1. In testing, we found that the classifier based on multi-band KiDS mock data (bottom) performed better than the classifier based on single-band Euclid mock data (top). }
  \label{roc}
 \end{figure}

\subsection{Data augmentation}

Machine learning methods often include a data augmentation step in order to increase the size of the training set (see e.g., \citealt{simard2003best, yaeger1997effective}). This may be necessary when the training set is sparse, or when the technique requires a very large sample of training data in order to generate a successful model - as can be the case for CNN. The practice of using label-preserving transformations to augment the training data was investigated during our application of SVM to the lens-finding problem. 

Beginning with the original training set, we aim to exploit invariances within the data by applying the following affine transformation to each image: rotation, inversion, shifting and scaling, as per the work of  \cite{2015MNRAS.450.1441D}. We also apply a colour perturbation as detailed by \cite{NIPS2012_4824}. This randomised process was applied to each image 10 times, ultimately increasing the training dataset by one order of magnitude. 

Concerns about the vulnerability of this step to overfitting were raised when it was observed that classification performance could vary significantly depending on how the augmented training data were used. During cross-validation, test results score significantly higher if the training and test data both include iterations of the same original images,  than when all 10 iterations of a single original image are  kept together. In the case of the former, it was concluded that the newly-generated images were not unique enough to avoid the modelling of random noise rather than the intended classification features. In the latter case, overfitting should not occur, but a lack of uniqueness in the training set could lead to bias and underfitting - and indeed cross-validation showed only a small improvement compared with the original data set. 

Subsequently, data augmentation has not been employed in our SVM application. However,  an augmentation procedure may yet become a useful step if a more nuanced method of data transformation were to be used.  For example, \cite{simard2003best} showed that CNN classification of handwritten digits could be improved by using elastic distortions in addition to affine transformations when augmenting the original training data. Such a technique is dependent on the inherent invariance of the data to elastic deformation  and maybe not suit the problem of lens classification, but is worth exploration.

% Lens finding challenge - where everything is on frastra
%

\section{Lens finding challenge}

The lens finding challenge (Metcalf et al. 2017, in preparation) was devised in order to compare the ability of competing techniques to find galaxy-galaxy lens systems in both the upcoming Euclid survey and in its ground-based counterpart, KiDS. The Euclid telescope will observe from space at visible and near-infrared frequencies. With a FOV of 15 000 square degrees, Euclid will image around 10 billion sources in order to use weak lensing to map the distribution of dark matter over time and constrain dark energy. Simulations of the survey predict an observable strong lens incidence of 1 in 3000 sources, amounting to $\sim$300000 lens systems within the whole survey \citep{2011arXiv1110.3193L}. KiDS \citep{2013ExA....35...25D} is an ongoing survey which uses four optical bands to map the weak lensing shear signal in 1500 square degrees of sky. Image quality improves on that of the Sloan Digital Sky Survey and observations have 2.5 magnitudes greater depth. 

Organised by the Bologna Lens Factory, the lens-finding challenge ran for ten weeks, during which time competitors could download mock catalogues containing labelled images with and without sources. Two datasets were simulated; a ``space'' dataset consisting of single-colour images with similar resolution to {\it Euclid}, and a ``ground'' dataset with resolution appropriate to a good terrestrial site, but with four colours ($U$, $G$, $R$, $I$). Each test dataset consisted of 100000 objects for blind classification, although a 20000-object catalogue was available in each case for training. The training catalogue was available throughout the challenge, but classification and submission of results from the test data, in the form of a confidence level for each object between 0 and 1, had to be achieved within a 48-hour period after download.

\subsection{Lens finding in the challenge dataset with SVM}
\label{calibrate}
We used the Gabor-SVM method to make submissions for both the ``space'' and ``ground'' challenge datasets. We converted SVM outputs to a confidence score by calibrating training scores. Since the relation of true positive rate to score tends to be a sigmoid distribution, a sigmoid function can be fit to the training data to convert the scores into probabilities. This so-called Platt-scaling \citep{Platt99probabilisticoutputs} is performed on held-out folds of the training data.  A successful calibration, when used in combination with a realistic simulated training set, has the advantage that each score will accurately reflect the probability of any one image being a lens or not; applied to real data, a score of 0.75 should mean a 0.75 chance of an image being a lens.

Fig.~\ref{scorescal} reports calibrated training scores for both datasets and illustrates the behaviour of the classifier. In both cases the method shows a high confidence in classifying images which contain sources but is less confident in classifying images without sources, particularly so for the single-band data. Areas under ROC curves of 0.80 and 0.93 were achieved for the final application to the challenge ``space'' and ``ground'' datasets, respectively. These scores are lower than those recorded during testing, particularly for the ``space''
 set. Here, a lack of generalisation could be a problem, despite the precautions taken to avoid overfitting. Alternatively, the use of different light-cones between mock training and mock challenge datasets may have produced significantly different Einstein radii distributions. The ``ground'' set score is not significantly lower than achieved during testing.  
 
For each dataset, training and application of the SVM to the challenge data totalled under 24 hours using a modest PC. The ``ground'' challenge results emphasised the ability of the Gabor-SVM method to maintain a high purity rate when classification scores are ordered from highest to lowest, with the method winning in the category for the highest number of correctly classified positive samples before encountering any false positives (Metcalf et al. 2017, in preparation).

\begin{figure}
  \centering
  \begin{tabular}{c}
  \includegraphics[width=1\linewidth]{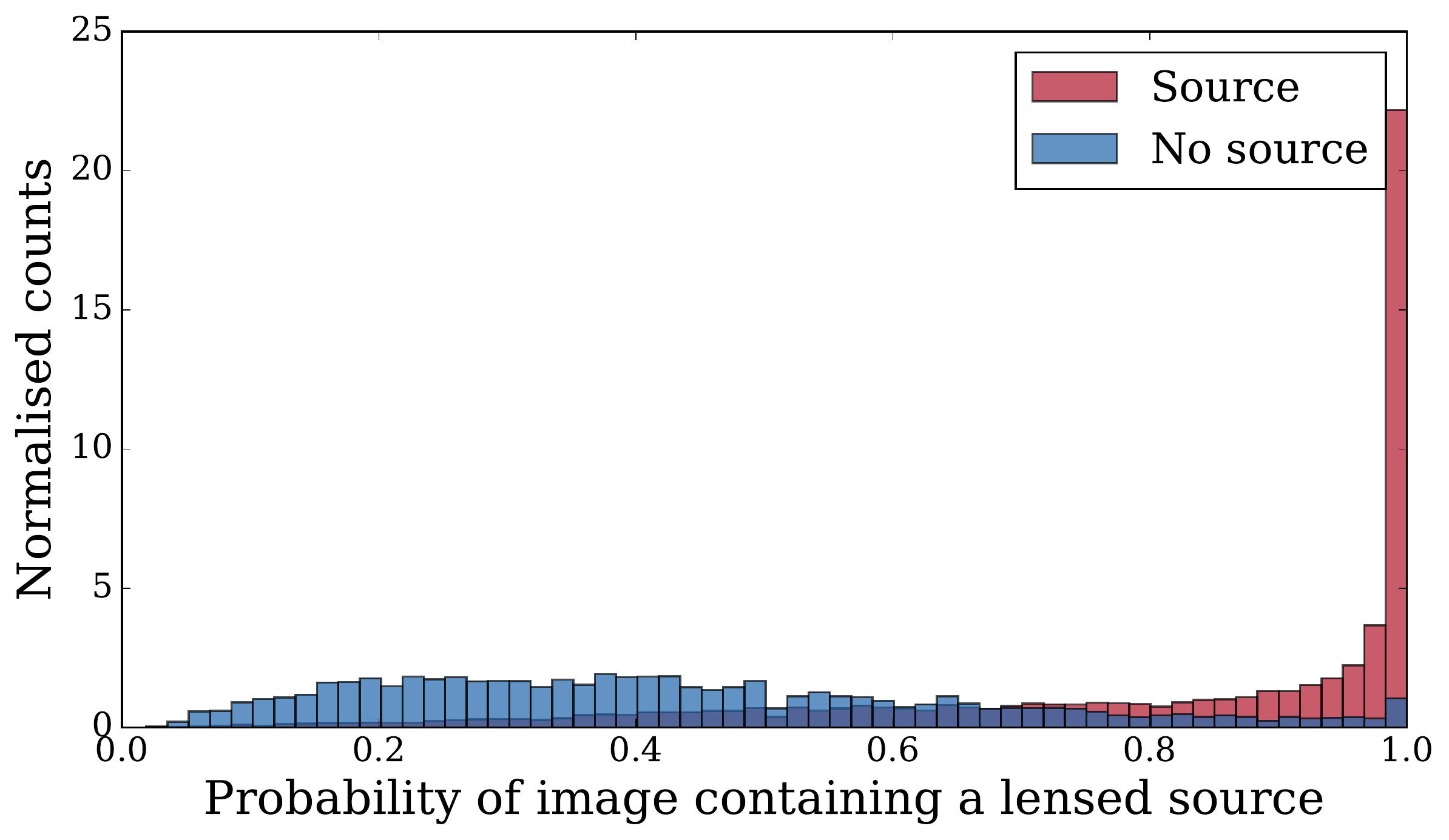}\\
  \includegraphics[width=1\linewidth]{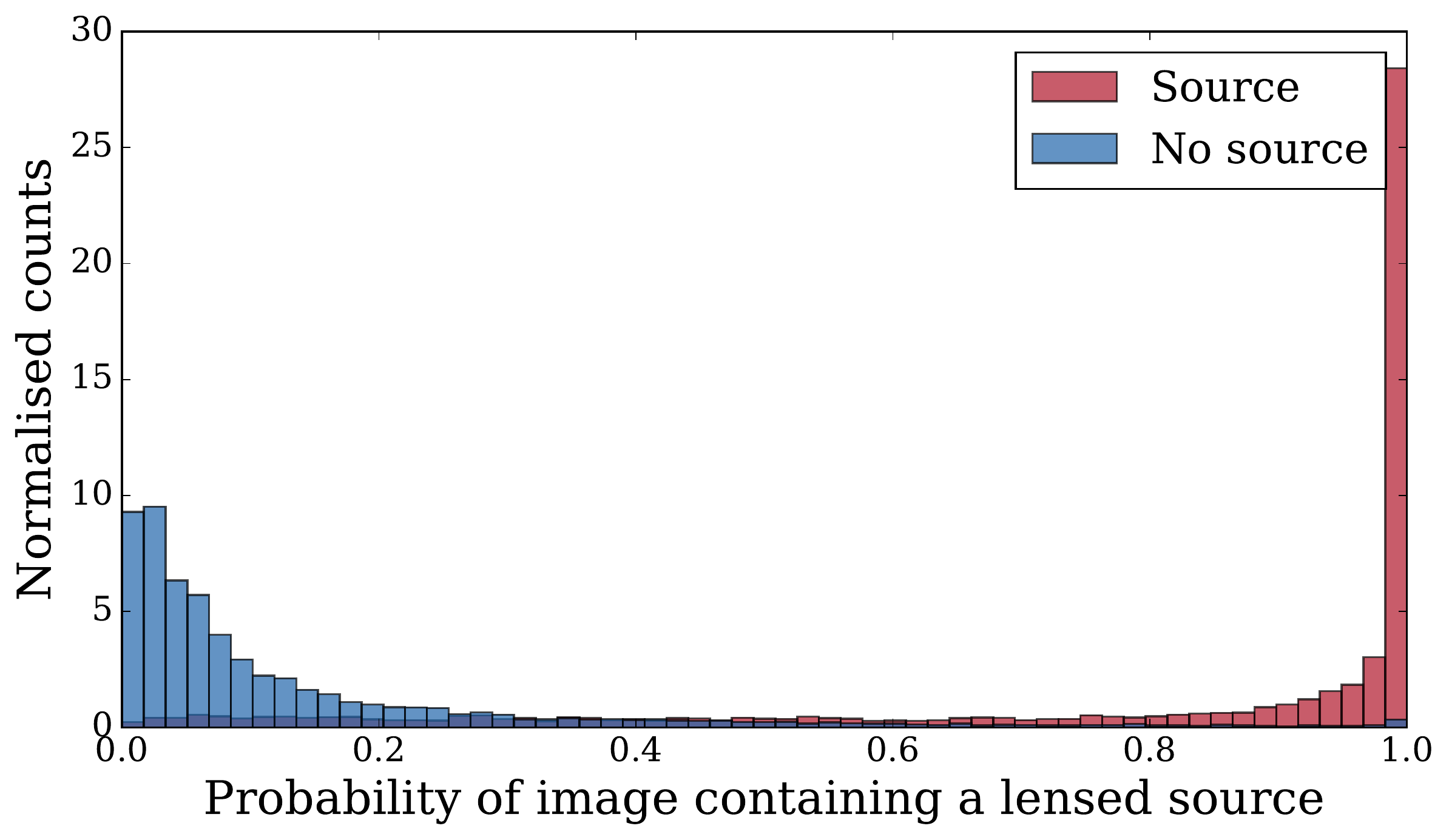}\\
  \end{tabular}
  \caption{ Calibrated training scores from the Lens Finding Challenge, for the single-band ``space'' mock dataset (top) and for the four band ground-based dataset (bottom). For the space-based set, the Gabor-SVM finder shows a high level of confidence in classifying images which contain a source, but is  less confident about images without a source. For the ground-based set, the method has improved confidence in classifying images without sources. Colour information obtained using multi-band observations appears to be important.}
  \label{scorescal}
 \end{figure}

\subsection{Lens finding in the challenge dataset with visual inspection}
\label{bigeye}
Future lens surveys with instruments such as {\it Euclid} will contain hundreds of millions or billions of individual objects, far too many for human inspection. However, the challenge datasets are still within the capabilities of human classifiers \citep{2008MNRAS.389.1311J,2015ARA&A..53..247M,2016MNRAS.455.1191M}. We therefore undertook inspection by eye of the 100000 objects in each of the ``space'' and ``ground'' categories.

To achieve this inspection within the allowed 48 hours, {\sc bigeye}, a Python tool\footnote{Available on {\tt https://www.github.com/nealjackson/bigeye}. The program also uses the {\tt img\_scale} routine, written by Min-Su Shin and available on {\tt http://dept.astro.lsa.umich.edu/$\sim$msshin/science/code/ Python\_fits\_image/img\_scale.py}.} was written to allow efficient processing of large numbers of images. The most important part of the design is the colour scaling of these images, in order to allow reliable identification both of high surface brightness lensed features, and also of faint arcs buried in diffuse emission from the lensing galaxy. This was achieved by considering pixel values only in the inner ninth of each image, in order to avoid being affected by bright nearby objects. Within this inner region, pixel values were arranged in ascending order and the bright burnout level of the colour table chosen as the pixel value at a fixed percentile within this ordered list. In practice, percentile values between 95 and 98 were useful, and the exact value in each case was determined by optimization (by trial and error) of areas under the ROC curve in the training set. An additional lower limit, of approximately 10 times the noise level, was imposed on this burnout level to cope with very faint images, where the noise in the image became unduly distracting from the task of pattern recognition. Images were then scaled between zero and the burnout level using a square-root scaling, converted to RGB images if the input data contained three or more colours, and assembled on a grid with a user-definable number of images per grid, and written to binary data files using Numerical Python for fast access by the observer. For this challenge, we used only $G$, $R$ and $I$ bands for three-colour reproduction, due to the much greater noise in the $U$ images.

\begin{figure}
\includegraphics[width=9cm]{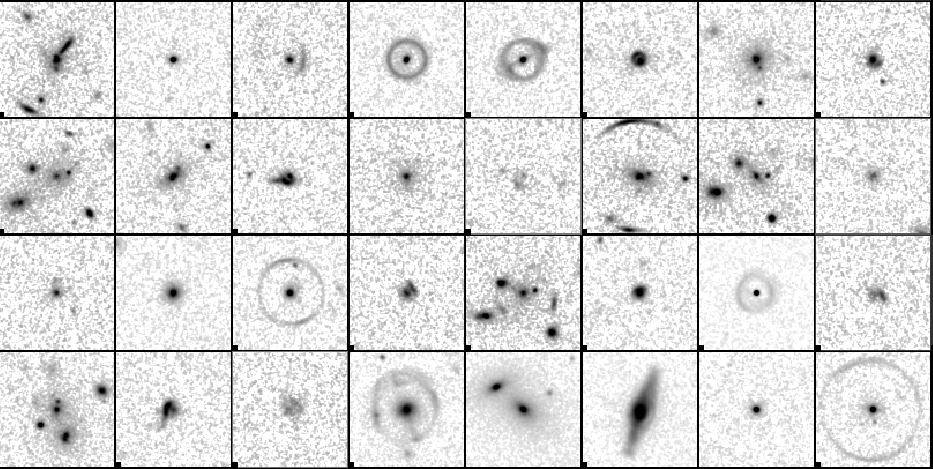}
\caption{Some of the ``space'' challenge images examined by eye. The image shows a typical 8$\times$4 grid made from the challenge training set. In this figure, true lenses are indicated by small black squares in the bottom left hand corner of each image.}
\label{bigeye_screen}
\end{figure}

The script allows the user to interact with the gridded images using the {\sc widgets} library within {\sc matplotlib}. Positions and keys of cursor hits are recorded in a file which contains a record of the image name, together with the corresponding cursor keys. For this purpose we used numbers from 0 to 4 to represent degrees of confidence ranging from 0 (no evidence of lensing) to 4 (certain lens); no cursor record was automatically regarded as equivalent to 0 on this scale. It is also possible to change the colour-scale interactively on this plot by altering the burnout level downwards, in order to see faint arc emission in galaxies.

Inspection of the images was done by two of the authors, NJ and AT, over the 48-hour period for each of the two challenge datasets (``space'' and ``ground''). A rate of $\sim$5000 images per hour, with 70000 inspected images (NJ) and $\sim$2500 images per hour, with 30000 images (AT) was achieved; we can thus evaluate any possible advantage of more careful inspection, as well as comparing the results to machine learning outputs. In practice, the area under the ROC curve was almost indistinguishable for both datasets, with values of 0.800 and 0.812 obtained for the ``space'' challenge, respectively, and 0.891 and 0.884 for the ``ground'' challenge.

\begin{table}
\begin{tabular}{lccccc}
 & 0 & 1 & 2 & 3 & 4 \\ \hline
 Space, lenses & 13293 & 8370 & 4281 & 3308 & 10723 \\
 Space, non-lenses & 53384 & 6321 & 251 & 51 & 18 \\
 Ground, lenses & 8374 & 7288 & 4637 & 4000 & 25636 \\
 Ground, non-lenses & 42159 & 5730 & 1172 & 539 & 465 \\ \hline
\end{tabular}
\caption{Classification statistics for the eyeball examination of the ``space'' and ``ground'' lens samples. Lenses are classified from 0 (no evidence of lensing) to 4 (virtually certain lens). Results from both evaluators have been combined. A perfect classification would have a classification of 4 for all lenses and 0 for all non-lenses in each case.}
\label{bigeye_stats}
\end{table}

In Table~\ref{bigeye_stats} we show the detailed results of the by-eye inspections. In general, the false positive rate corresponding to the highest-confidence lens identification of images which are not lenses is quite low - generally between 0.1\% and 0.3\%. The much higher rate of false positives in the ground-based data ($\sim$2\%) can be traced to an image which occurs repeatedly throughout the simulated dataset and consists of two very similar objects arranged symmetrically about the central galaxy in a possible lens configuration. One observer classified these objects as generally with a classification of 2 (thereby obtaining a 0.3\% false positive rate), the other generally with a classification of 4.

% Where stuff is on frastra:
%
% Ground-based dataset used for eyeballing: catalogue 5
% Copy of files on /mirror1/scratch/njj/ground_data
% Bigeye submission files and truth data: /mirror1/scratch/njj/ground_files
%
% Ground-based dataset used for original SVM:
%
% Space-based dataset used for both: catalogue 1
% Copy of data files on /mirror2/scratch/challenge_space
% Bigeye submission files and truth: /mirror1/scratch/njj/space_files
%

\subsection{Comparison of SVM and visual inspection}
\label{compare}

\begin{figure*}
  \centering
  \begin{tabular}{c}
  \includegraphics[width=18cm]{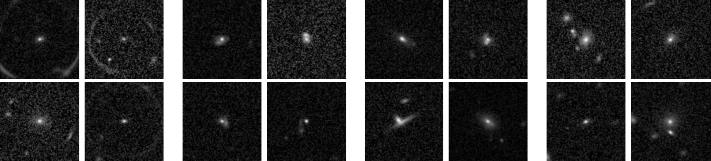}\\
  \\
  \includegraphics[width=18cm]{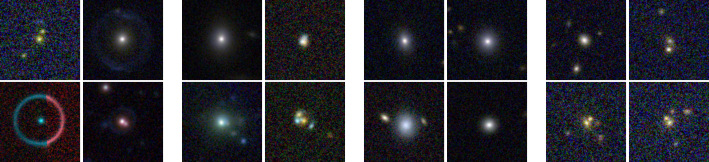}\\
  \end{tabular}
  \caption{Comparison of classification disagreements made between the SVM method and the visual inspection method using (top two rows) the same space-based mock lens system candidates and (bottom two rows) ground-based candidates. In each case the left-hand panel (four images) shows the SVM false negatives with the lowest assigned probability, for which the eyeball search gives a firm lens identification. The second panel shows the visual inspection false negatives with the highest (most likely lens) eyeball classification. The third panel shows the SVM false positives with the highest lens probability, which are rejected by eye. The final panel shows the worst visual-inspection false positives, which have low lens probabilities using SVM. }
  \label{inspectspace}
  \end{figure*}

The large sample of challenge mock images allows for a useful study of the bias of competing methods. Of particular interest is the comparison of the performance of automated methods with the performance of visual inspection, which has historically been seen as the most reliable way of finding lens candidates for follow-up observation. Recent lens finding applications also detail the use of the eye as a final classification pass after application of automated methods to large surveys, and the SPACE WARPS project is performed exclusively by eye. In Fig.~\ref{inspectspace} we examine samples which received contradictory classifications by the Gabor-SVM and visual inspection methods. For both the ``space'' and ``ground'' challenge sets we identify the images which received the highest scores from one method but the lowest scores from the other method and divide these groups into those containing a mock lens-system (positive samples) and those which do not contain a lens-system (negative samples).

We find that the Gabor-SVM method misses some positive samples which to the eye are very obviously lens systems. Part of this failure is due to the large radii of some lens images cut off by the polar transformation, which extends only to the radius of the length of half the image width. This can easily be remedied in future applications. The method also misses some faint rings and some rings corrupted by masks.  The population of lenses identified by the Gabor-SVM method but missed by eye is dominated by small-radii lens systems.  Observations of this subset may be difficult to observe and model - higher resolutions would be required for detailed imaging of both the lensing pattern to study the gravitational potential, and for the lensing galaxy itself, so that information from stellar velocities can be used to break the so-called mass-sheet degeneracy \citep{2013A&A...559A..37S,2014A&A...568L...2S}. Yet, as simulations have shown, the radii missed by eye form the bulk of lensing systems - with the peak in radii distribution expected to occur at around 1 arcsec \citep{2015ApJ...811...20C}. Therefore, if we are to detect the predicted lensing rate of 1 in $>$1000 sources within the Euclid dataset, detecting the population of small systems will be important. Furthermore, by ensuring that the parameter space is kept as open as possible, we will be more likely to find exotic compound lenses (e.g. \citealt{0004-637X-752-2-163}) which could prove more robust to degeneracy problems \citep{2012MNRAS.424.2864C}. Fig.~\ref{ein_rad} illustrates the performance of the respective methods with increasing maximum Einstein radius. The Gabor-SVM method displays the potential to recover systems outside the historical constraints of visual inspection, potentially probing populations which have been discounted in previous detection criteria. However, since visual inspection as a follow-up method would be redundant in such cases, a high confidence in the automated method would be required. 

Of the negative samples, the eye is successful in ruling out tangential components or diffuse, blue, face-on spiral galaxies that the Gabor-SVM falsely classifies as positive. These objects are likely to be a significant problem for real survey lens-finding where such objects will feature much more commonly than in the challenge mock datasets. Future automated methods will benefit from training samples which include high numbers of such objects in order for a classifier to successfully learn these distinctions. Human-classified false-positives which the Gabor-SVM method discounts are largely those which mimic multi-point source lenses.

%\begin{figure}
%  \centering
%  \includegraphics[width=1\linewidth]{ein_dist.png}
%  \caption{}
%  \label{ein_dist}
  
%  \end{figure}

\begin{figure}
  \centering
  \includegraphics[width=1\linewidth]{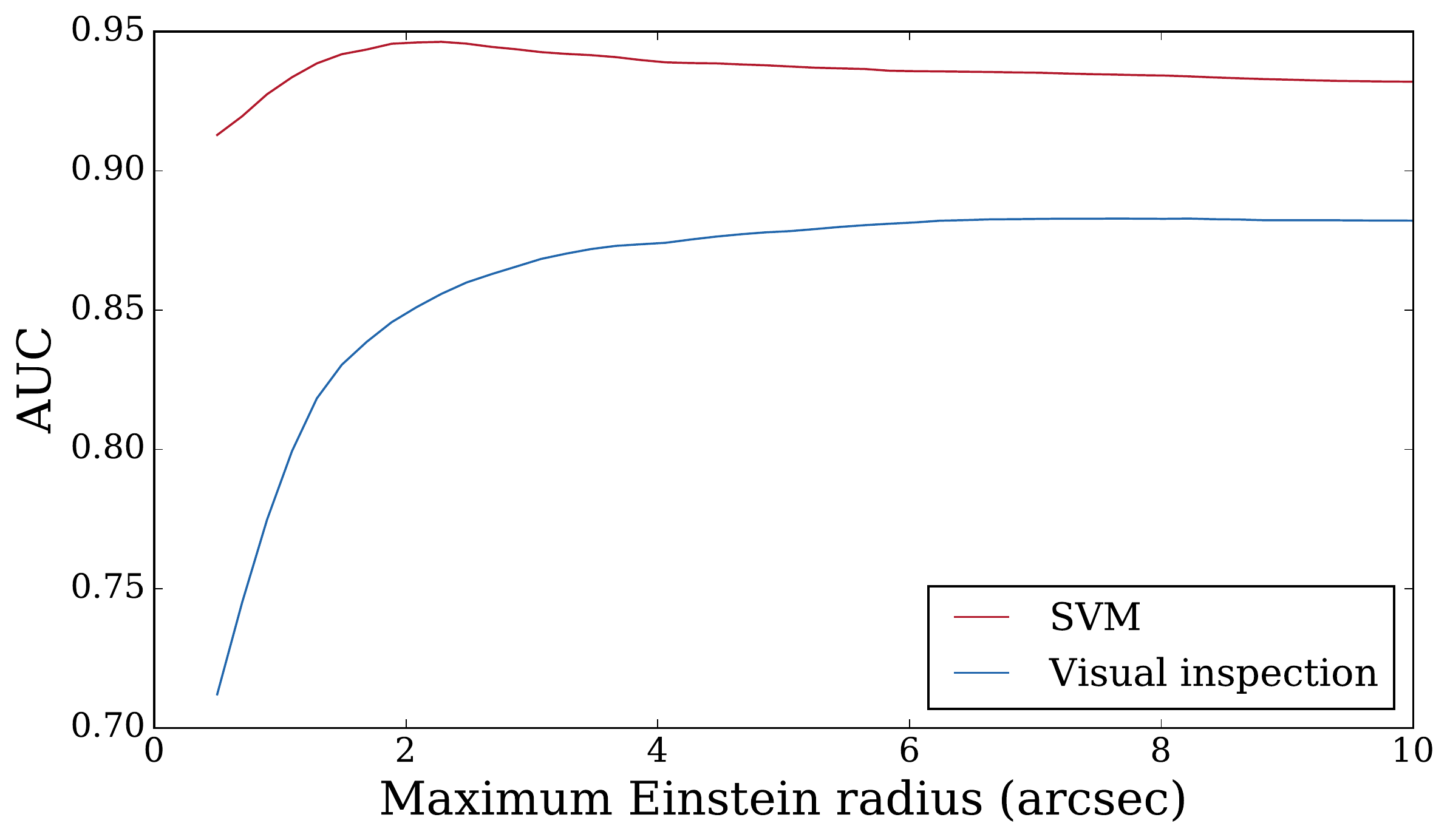}
  \caption{AUC scores for both the Gabor-SVM and visual inspection methods are plotted against increasing maximum Einstein radii of lensing sources. Visual inspection performs better at the higher radii, where contaminants such as spiral galaxies, which were not well represented in the training data, become a problem for the SVM. At low radii, where the bulk of the lensing population is expected to be found, the SVM method is more successful. }
  \label{ein_rad}
  
  \end{figure}
%plots of distribution of radii and snr, and percentages recovered by svm/eye. How small do the blf lenses go? Is this smaller than colletts lenses? How do they compare to collett's detection criteria? tautological argument? Collett says that the lens radii minimum is limited by the characteristics of the ringfinder detection method, or alternatively the human inspection method, where only lenses meeting three? morphological criteria plus a snr cut are counted. The SVM method displays the potential to recover systems outside those constraints, with rather different selection effects from the former methods. It would be interesting to see whether automated method could detect yet smaller Einstein radii lenses.

%look at svm results per einstein bin and snr bin and see how many would be recovered from cfht, where Gavazzi found 220. Would it find more - and smaller - lenses - what is the seeing distribution of these?

\section{Application to real data}

After using the ground-based mock data set from the Bologna Lens Factory to obtain a trained model, we have applied our  classifier to real ground-based data using the same bands from the KiDS survey. We downloaded images from the third KiDS data release, made publicly available by the European Southern Observatory\footnote{{\tt http://kids.strw.leidenuniv.nl/DR3}} \citep{2015A&A...582A..62D}. Data are provided in the form of {\sc fits} files and accompanying source catalogues compiled using {\sc SExtractor}. Image resolution is 0\farcs2 per pixel and limiting magnitudes are  $u=24.8$, $g=25.4$, $r=25.2$ and $i=24.2$ (for $5\sigma$ 2\arcsec AB), with exposure times chosen to provide a sample with median redshift 0.7. The whole survey will ultimately cover 750 square degrees each in the Northern and Southern skies.
In order to assemble a sample of objects for classification, we used the accompanying catalogues to select all objects which matched several criteria. To remove stars from the sample we selected objects for which the catalogue parameter CLASS\_STAR $<0.8$. Spiral galaxies, relatively blue in colour and featuring arc-like extensions, share a similar morphology to lens patterns. Since spirals were not common in our training set our model will lack the ability to distinguish between lensed arcs and  contaminating objects such a spiral arms.  To reduce the occurrence of spiral galaxies we made a colour cut of $g-i>1$, an ellipticity cut of $A/B<3$ and a magnitude cut of $r<22$. Our remaining sample totalled 1011403 objects to which our finder was applied. 

The distribution of classification scores obtained from the sample shows an overall bias towards positive classification (see Fig.~\ref{histscores}); since the scores have been calibrated as probabilities (see Section~\ref{calibrate}), the sum of all scores should represent the expected observed lensing incidence of the order 0.001. In practice, the scores from this sample sum to around 0.5. The discrepancy highlights the need for training samples to include more contaminating objects such as spiral galaxies and multiple objects, which, here, have contributed to an apparent high incidence of strong lens systems.

\begin{figure}
  \centering
  \includegraphics[width=1\linewidth]{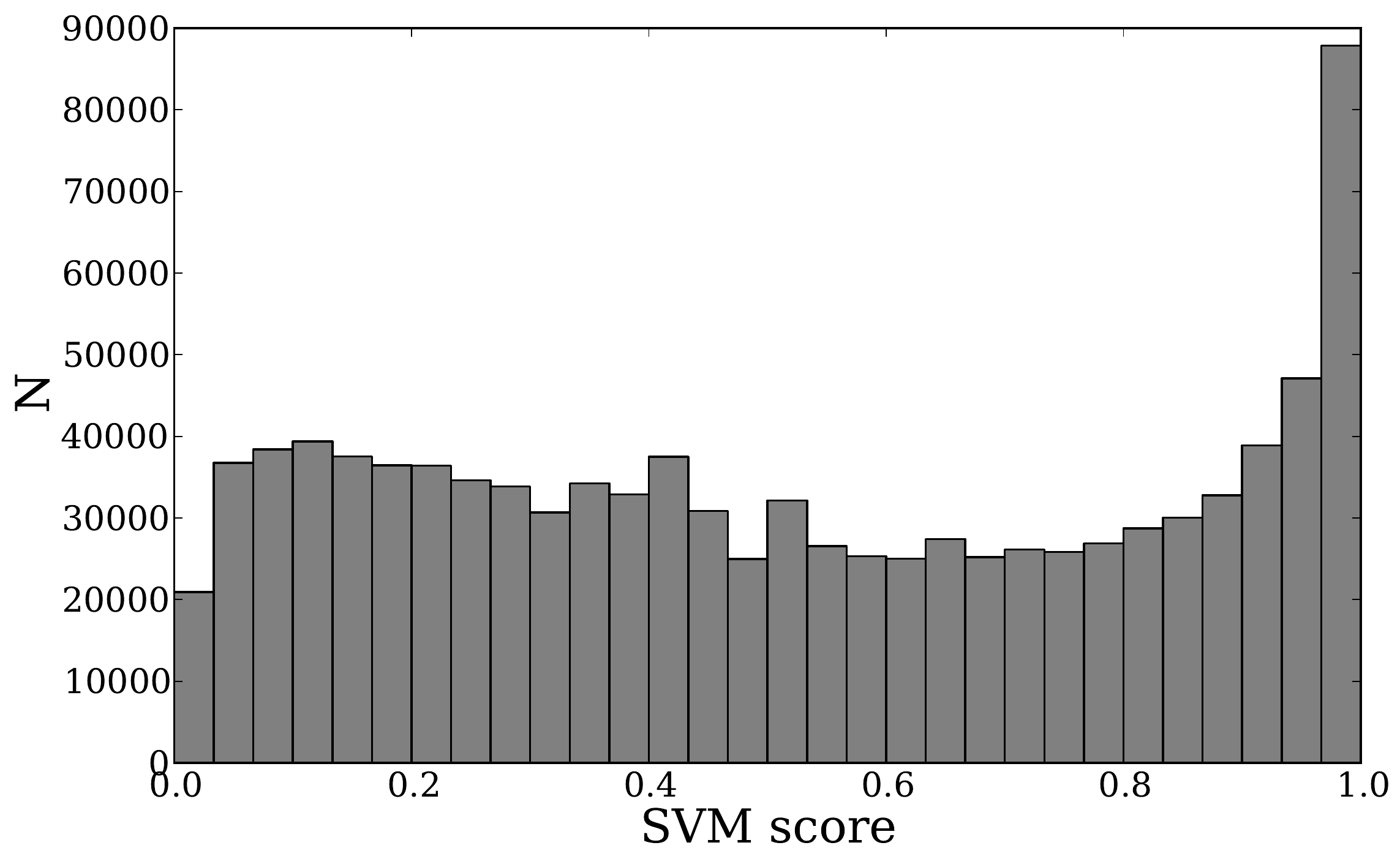}
  \caption{Classification scores expressed as probabilities for all objects selected from the KiDS catalogue. The overall distribution is greatly skewed towards the high end and the sum of all probabilities predicts that approximately half of all objects  should contain lensed sources. Since the probability calibration was previously able to make successful predictions of total lensed source counts  for unseen challenge data, we assume that the exaggeration here is due to the real KiDS images containing many more contaminating features than the mock dataset used for calibration. Improvement to both the classification performance and the calibration of scores should be seen  by using a training set which contains a more representative distribution of contaminants such as spiral galaxies and multiple background sources. }
  \label{histscores}
  
  \end{figure}

Since the score calibration cannot be used to provide direct probabilities in this case, for our final classification step we simply ranked images by scores and used our multi-image  display tool {\sc bigeye} (Section~\ref{bigeye}) to visually  inspect the highest-scoring 10000 images. At a rate of ~1\% of the total sample, this amount of inspection  will be unfeasible for larger surveys containing billions of sources. Within this set we found a high proportion of early-type spirals - blue arms surrounding whiter cores - confused with the arcs of a lens system, as expected. We saved 1000 promising objects for further inspection, of which we expect polar ring galaxies and spirals - particularly type 2a spirals, where a relatively bright nucleus and tightly wound arms is often difficult to distinguish from a lens by eye - to form a proportion. In Appendix A we present 213 objects found in the KiDS third data release which range from possibly to very likely lenses. Fig.~\ref{lenscans} displays in greater detail five of the lens candidates we have found. The finder shows the ability to detect lens systems of varying morphology, with two very promising four-object lensed quasar candidates and several possible double-object lensed quasars residing among the sample of ring- and arc-like objects.

  \begin{figure}
  \centering
  \includegraphics[width=1\linewidth]{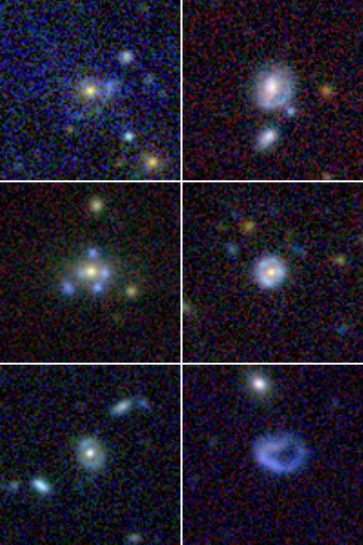}
  \caption{A selection of five of our lens candidates found in the KiDS third data release (see Appendix A). The candidates were  obtained after visual inspection of the 10000 highest scoring objects from application of our Gabor-SVM lens finder to a total of 1011403 catalogue entries. The bottom right image shows a ``smoke ring'' galaxy, possibly the result of a collision with its compact neighbour.}
  \label{lenscans}
  
  \end{figure}

Among the highest 10000 scoring candidates we found a ``smoke ring'' galaxy, characterised by a ring of star-forming material with no obvious nucleus (see Fig.~\ref{lenscans}). Empty rings are thought to form when a compact  galaxy collides with a large gas-rich system, producing a density wave which radiates material outwards and triggers the ring of star-formation (see \citealt{appleton1996collisional} for a review of collisional galaxies and \citealt{2009ApJS..181..572M} for a more recently collated atlas). The object we have found is not perfectly elliptical and its apparent wholeness could also be the result of the projection of a chance alignment of distorted spherical arms. A region of redder material embedded within the blue hints at the remains of a nucleus. An  elliptical companion is located 8 arcseconds away on the sky. The KiDS catalogue provides photometric redshifts calculated using  BPZ: a Bayesian redshift calculator which incorporates priors in the form of object type and  brightness to obtain a likelihood function (see \citealt{2000ApJ...536..571B} for application of BPZ to HST data). The catalogue lists two entries for the ring, which has been detected as separate components by SExtractor during the catalogue compilation. The redshifts of these two entries are 0.43 and 1.14. ``ODDS'' of the values being correct to within $\Delta z=0.1$ are 0.806 and 0.441. The redshift of the nearby elliptical galaxy is reported as 0.45, with an ``ODDS'' value of 0.999.  Given the problems with redshift measurement of the ring, it is possible that its redshift and the redshift of the elliptical object are similar enough to indicate an interacting pair and that the empty ring is a result of a collision with the compact elliptical. The system could provide a useful study of the merger stages of galaxy evolution and of dark matter dynamics of galaxy collisions. The discovery also highlights the potential ability of the Gabor-SVM method to discover ``dark lenses'' and other exotic objects.

Our comparison of visual inspection methods with the Gabor-SVM technique (Section~\ref{compare}) showed that there will be a significant number of lens systems which may be detected by machine but not by eye. Knowing this, our final step of visual inspection of the KiDS data is unsatisfactory in terms of completeness of the lensed population, but this is unavoidable at this stage.  With evermore realistic training sets and further developments to the method, we would hope that the machine would be able ultimately to identify with a high confidence those systems which are undetectable by eye. 

\subsection{Comparison with CNN finder}

\cite{2017arXiv170207675P} have previously applied a convolutional neural network (CNN) to the same survey data to find 56 new lens candidates, 22 of which show morphological consistency with Einstein radii expected from lens masses estimated using velocity dispersions. The CNN finder was trained using an augmented sample of 
luminous red galaxies (LRG) and contaminants from the KiDS dataset, along with simulated lensed sources. It was  applied to a total of 21789 KiDS LRGs obtained using a colour-magnitude selection. Of the candidates presented, we find five objects which obtain a score greater than 0.9 using the Gabor-SVM finder (see Fig.~\ref{cnnlenses} for examples). The CNN finder was developed and trained to find LRG lens systems specifically, whereas the SVM-Gabor finder uses a different and broader subset of training galaxy types. A more direct comparison using the same set of training data to classify the same set of survey objects would be interesting.

%BPZ 0.43... 0.449 80 million pc apart radially 52200 tangentially 1,815,000,000pc  1,896,000,000pc - size of a cluster,too far?

%galaxy zoo paper about colours \citep{2013MNRAS.432..359T} does this consider blue ellipticals as possible lenses? red cut off is a fit of $g-r=0.73 -0.02(M_r+20)$ for sdss absolute mag \citep{MNR:MNR16503}

 % \begin{figure}
 % \centering
 % \includegraphics[width=1\linewidth]{cnnvssvmplot.png}
 % \caption{score distributions - leave this out??}
 % \label{cnnvssvm}
  
  %\end{figure}

  \begin{figure}
  \centering
  \includegraphics[width=1\linewidth]{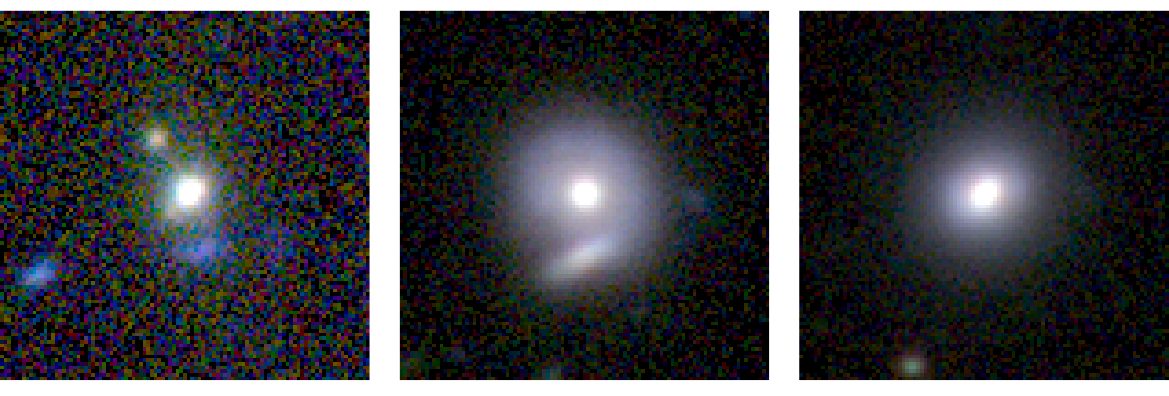}
  \caption{Some of the objects classified as lens candidates by the CNN method of \protect\cite{2017arXiv170207675P}, which received a high score ($>0.9$) from our method.}
  \label{cnnlenses}
  
  \end{figure}

\section{Discussion}

We have developed an automatic classifier which uses a novel application of machine learning and image processing methods to identify galaxy-galaxy scale strong gravitational lens systems in large datasets. We trained and tested a support vector machine using two mock data sets produced by the Bologna Lens Factory. One set contained images representing the visible (VIS) band of the upcoming Euclid survey, while the other contained four image bands representing the Kilo Degree Survey.  Input to the SVM is in the form of statistics derived from the application of a Gabor filterbank to polar-transformed postage stamp images centred on sources. We paid particular attention to both the selection of training features and to the regularisation of the SVM kernel so that the model will generalise well to new datasets.   The method has shown promise both in testing during the Lens Finding Challenge, where it achieved first place in the ``ground'' challenge in the category of false positive rejection (Metcalf et.al., in preparation), and in application to real data from the third release of the Kilo Degree Survey, where we have found a large selection of possible lens candidates suitable for follow-up study. Of particular interest is the ability of the finder to retain a high level of purity; when applied to future survey catalogues containing billions of sources, the ability to avoid false positives will be essential. 

Obtaining a good set of input features for the SVM is not a trivial task. The use of Gabor filters to provide a generalised approach to feature extraction was inspired by current advances in the field of computer vision, and has proved promising. Filter convolution was the most intensive step of image processing and the time dependence on filter kernel size prevented extension into the low frequency filter domain, where signal appears to be important. Implementation using a more powerful machine would allow for this and should further improve classification performance. There is also an undesirable dependence of the signal of the Gabor filter on the average pixel value of an image due to a non-zero DC response. This could be removed by using Log-Gabor filters which use a logarithmic frequency response to avoid this problem (see \citealt{zhitao}).  Further investigation into filter choice and statistical representation is expected to improve the SVM technique. 

A  convolutional neural network has the advantage that the method  itself learns which features are of most importance for successful classification, allowing for the extraction of abstract feature sets which describe the problem well, without the need for human intervention. Several examples of successful CNN application to classification problems can be found in the literature. However, in the case of lens-finding we are concerned with classifying rare objects, and the ability of the SVM technique to reject false positives becomes particularly important. The results of the Lens Finding Challenge indicate the potential success of hybrid methods, where CNN are used to provide features for SVM classification. The purity  advantage of the SVM may be due to the deterministic nature of the classification model obtained by maximising the margin between classes: a well-regularised model will be unlikely to permit false classifications near the extremes of each class. Neural networks, on the other hand, use a probabilistic combination of hypotheses,  which can in some cases provide robust generalisation to new datasets but also possesses inherent randomness, possibly leading to a small but significant distribution of falsely classified samples at the extremes (see \citealt{2016arXiv160202389F} for an investigation into the robustness of deep learning methods).

We predict significant improvements to the performance of the Gabor-SVM finder with the provision of increasingly realistic training samples. Training images containing representative incidences of polar ring and spiral galaxies, as well as general background sources, will further reduce the number of false positives. We note the importance of colour information for the success of the classifier. Results in testing showed a better performance on multi-band rather than single-band mock data sets. We expect improvements in the classification of  datasets representing the Euclid survey with addition of mock data samples representing the three low-resolution bands (J, H and Y) of the Near-Infrared Spectrometry and Photometry instrument (NISP) to the mock VIS band data.

The comparison of results using the Gabor-SVM classifier versus visual inspection has highlighted the significant statistical difference between samples of lens systems found by the respective methods. Simulations predict that systems with small ($<$1\arcsec) Einstein radii form the bulk of lensing incidents on the sky. Systems of this size are found by our machine learning method but are difficult to detect by eye without very careful subtraction of the potential lensing galaxy. Without a high level of  detection confidence, follow-up observations would be difficult. With increasingly representative training data and further development to the method, our classifier shows potential to confidently classify this population of objects. This could significantly increase the overall potential number of lens systems available for statistical investigation of galaxy evolution, structure formation and cosmology. 

Our method can be easily adapted to other large survey instruments such as the LSST and, at radio frequencies,  the SKA. Further modification could be made for radio interferometer survey data by using a set of features derived from the u-v plane as input to the SVM.

\section*{Acknowledgements}

We thank the KiDS collaboration for providing data in advance of publication. PH acknowledges receipt of an STFC studentship, and AT acknowledges receipt of an STFC postdoctoral fellowship during the course of this work.  We acknowledge support during the preparation of this work from the International Space Science Institute (ISSI), Berne, Switzerland, in the form of support for meetings of the collaboration ``Strong Gravitational Lensing with Current and Future Space Observations'', P.I. J-P. Kneib.
%%%%%%%%%%%%%%%%%%%%%%%%%%%%%%%%%%%%%%%%%%%%%%%%%%



\bibliographystyle{mnras}
\bibliography{phd,extra} % if your bibtex file is called example.bib

%%%%%%%%%%%%%%%%%%%%%%%%%%%%%%%%%%%%%%%%%%%%%%%%%%

%%%%%%%%%%%%%%%%% APPENDICES %%%%%%%%%%%%%%%%%%%%%

\appendix

\section{Lens systems candidates from the KiDS survey}
Here we present a selection of potential lens candidates found by application of the Gabor-SVM finder to the third data release from the Kilo Degree Survey.

\begin{figure*}
\includegraphics[width=15cm]{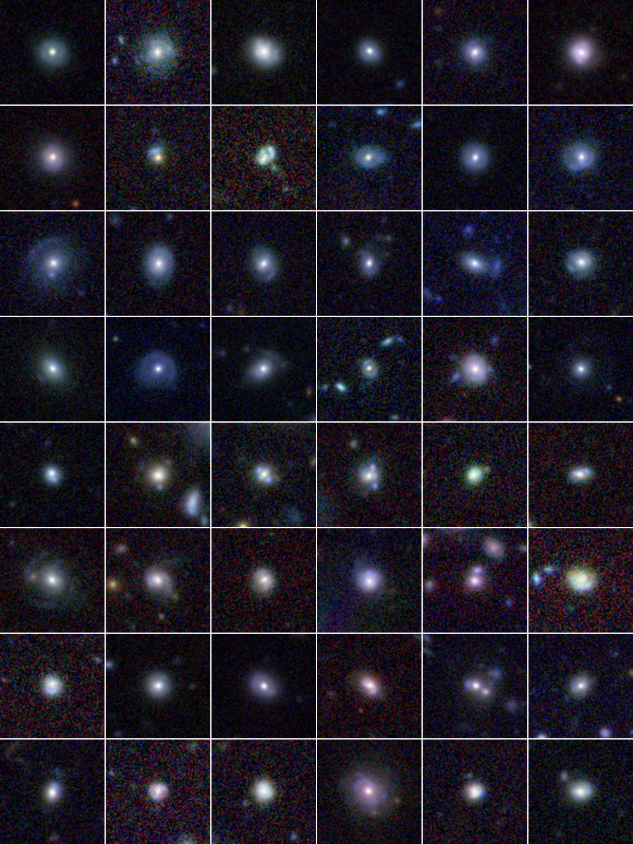}
\caption{A selection of the highest scoring lens candidates from application of the Gabor-SVM finder to the third data release from the Kilo Degree Survey. Images are 100 pixels wide with a resolution of 0\farcs2 per pixel.}
\label{candidates3}
\end{figure*}

\begin{figure*}
\includegraphics[width=15cm]{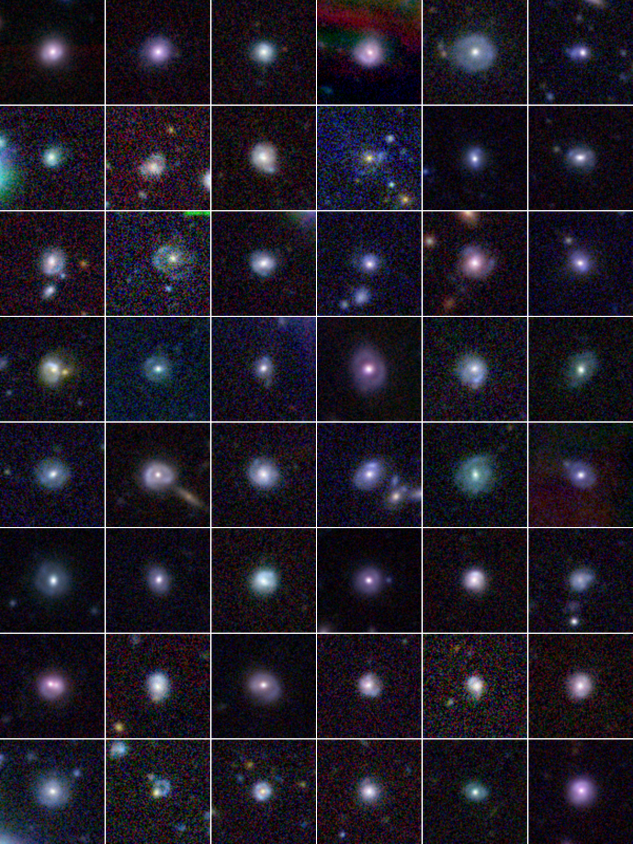}
\caption{A selection of the highest scoring lens candidates from application of the Gabor-SVM finder to the third data release from the Kilo Degree Survey. Images are 100 pixels wide with a resolution of 0\farcs2 per pixel.}
\label{candidates4}
\end{figure*}

\begin{figure*}
\includegraphics[width=15cm]{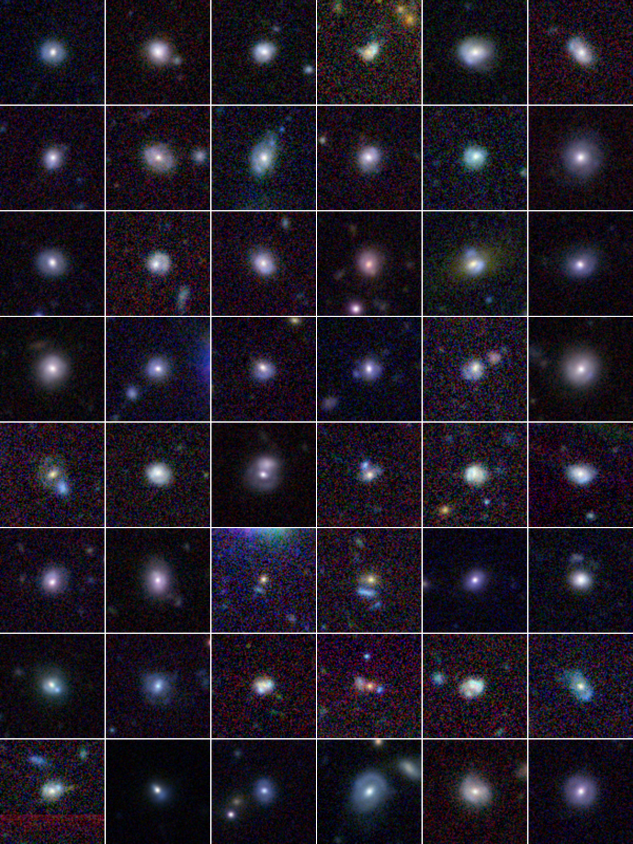}
\caption{A selection of the highest scoring lens candidates from application of the Gabor-SVM finder to the third data release from the Kilo Degree Survey. Images are 100 pixels wide with a resolution of 0\farcs2 per pixel.}
\label{candidates5}
\end{figure*}

\begin{figure*}
\includegraphics[width=15cm]{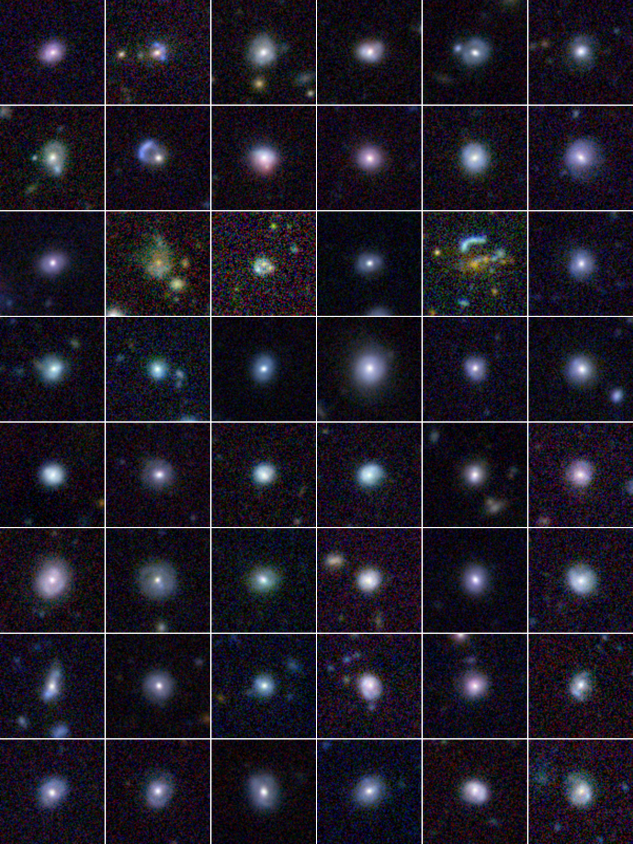}
\caption{A selection of the highest scoring lens candidates from application of the Gabor-SVM finder to the third data release from the Kilo Degree Survey. Images are 100 pixels wide with a resolution of 0\farcs2 per pixel.}
\label{candidates6}
\end{figure*}

\begin{figure*}
\includegraphics[width=15cm]{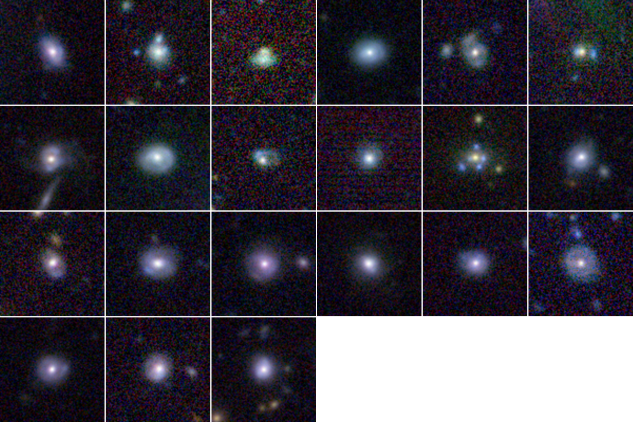}
\caption{A selection of the highest scoring lens candidates from application of the Gabor-SVM finder to the third data release from the Kilo Degree Survey. Images are 100 pixels wide with a resolution of 0\farcs2 per pixel.}
\label{candidates6}
\end{figure*}

%%%%%%%%%%%%%%%%%%%%%%%%%%%%%%%%%%%%%%%%%%%%%%%%%%

% Don't change these lines
\bsp	% typesetting comment
\label{lastpage}
\end{document}